\documentclass[preprint]{aastex}
\tighten

\input psfig

\def\kms{\,{\rm km\, s^{-1}} }
\def\kmsm{\,{\rm km\, s^{-1}\, Mpc^{-1}} }

\newcommand\scriptL{{\cal L}}

\begin{document}

\title{The Evolution and Structure of Early-type Field Galaxies:\\ 
A Combined Statistical Analysis of Gravitational Lenses}

\author{D. Rusin}
\affil{Department of Physics and Astronomy, University of
  Pennsylvania, 209 So. 33rd St., Philadelphia, PA, 19104-6396}

\author{C.S. Kochanek}
\affil{Department of Astronomy, The Ohio State University, 140 W. 18th
  Ave., Columbus, OH 43210}

\begin{abstract} We introduce a framework for simultaneously investigating
  the structure and luminosity evolution of early-type gravitational
  lens galaxies. The method is based on the fundamental plane, which
  we interpret using the aperture mass-radius relations derived from
  lensed image geometries. We apply this method to our previous sample
  of 22 lens galaxies with measured redshifts and excellent
  photometry. Modeling the population with a single mass profile and
  evolutionary history, we find that early-type galaxies are nearly
  isothermal (logarithmic density slope $n=2.06 \pm 0.17$, 68\% C.L.),
  and that their stars evolve at a rate of $d \log (M/L)_B/dz = -0.50
  \pm 0.19$ (68\% C.L.) in the rest frame $B$ band. For a Salpeter
  initial mass function and a concordance cosmology, this implies a
  mean star formation redshift of $\langle z_f \rangle > 1.5$ at 95\%
  confidence. While this model can neatly describe the mean properties
  of early-type galaxies, it is clear that the scatter of the lens
  sample is too large to be explained by observational uncertainties
  alone. We therefore consider statistical models in which the galaxy
  population is described by a distribution of star formation
  redshifts. We find that stars must form over a significant range of
  redshifts ($\Delta z_f > 1.7$, $68\%$ C.L.), which can extend as low
  as $z_f \sim 1$ for some acceptable models. However, the typical
  galaxy will still have an old stellar population ($\langle z_f
  \rangle > 1.5$).  The lens sample therefore favors early star
  formation in field ellipticals -- even if we make no {\em a priori}
  assumption regarding the shape of the mass distribution in lenses,
  and include the range of possible deviations from homology in the
  uncertainties.  Our evolution results call into question several
  recent claims that early-type galaxies in low-density environments
  have much younger stars than those in rich clusters.

\end{abstract}

\keywords{galaxies: elliptical and lenticular, cD -- galaxies:
evolution -- galaxies: structure -- gravitational lensing}

\section{Introduction}

Mapping the history of early-type (elliptical and lenticular; E/S0)
galaxies is essential for testing the modern picture of galaxy
formation. First, early-type galaxies are believed to result from the
mergers of later-type disk galaxies (e.g., White \& Rees 1978; Davis
et al.\ 1985; Kauffmann, White \& Guiderdoni 1993; Cole et al.\ 1994;
Kauffmann 1996), and therefore provide a unique probe of hierarchical
clustering in the cold dark matter (CDM) paradigm. Second, the
evolution of their stellar populations places constraints on the most
recent epoch of star formation, and hence, when mergers may have
occurred. Semi-analytic CDM models (e.g., Baugh, Cole \& Frenk 1996;
Kauffmann 1996; Kauffman \& Charlot 1998; Diaferio et al.\ 2001) have
investigated the evolutionary history of early-type galaxies, and find
significant environmental dependency. Specifically, they indicate that
E/S0s in high-density environments such as clusters should have
significantly older stellar populations than field ellipticals, many
of which should have formed in recent mergers. Current surveys are now
able to directly test these predictions by tracing the evolution of
early-type galaxies in various environments out to $z \sim 1$.

Ellipticals in rich clusters have been the most extensively studied,
with samples now approaching $z\sim 1.3$ (van Dokkum \& Stanford
2003). It is therefore not surprising that a broad consensus has been
reached regarding their properties. The evolution in mass-to-light
ratio inferred from the fundamental plane (FP) strongly indicates an
early mean star formation epoch ($\langle z_f \rangle \ga 2$). For
instance, the sample of van Dokkum et al.\ (1998) yields an evolution
rate in the $B$ band of $d \log (M/L)_B / dz = -0.49 \pm 0.05$ to
$z=0.83$, corresponding to a preferred formation redshift of $\langle
z_f \rangle \simeq 2.5$. The subsample at $z\sim 1.3$ is consistent
with an extrapolation of this result. Cluster galaxies are likely to
experience significant morphological evolution, and the mergers in
MS1054--03 (van Dokkum et al.\ 1999, 2000) appear to be direct
evidence for the hierarchical clustering model. Morphological
evolution can lead to a ``progenitor bias'', by which the early-type
galaxy population will appear systematically younger than its true age
(Kauffmann 1996; van Dokkum \& Franx 2001). However, applying the
maximal progenitor bias only changes the evolution rate to $d \log
(M/L)_B / dz = -0.56 \pm 0.05$ ($\langle z_f \rangle \simeq 2$; van
Dokkum \& Franx 2001). Because these and other evolution results (van
Dokkum \& Franx 1996; Jorgensen, Franx \& Kjaergaard 1996; Jorgensen
et al.\ 1999; Kelson et al.\ 1997, 2000; Wuyts et al.\ 2004) are
strongly supported by color and spectral analyses (Bower, Lucey \&
Ellis 1992; Ellis et al.\ 1997; Stanford, Eisenhardt \& Dickinson
1998; Ferreras, Charlot \& Silk 1999), there is little evidence to
indicate that early-type galaxies in present-day clusters are anything
but old, red and dead.

The history of early-type galaxies in low-density environments is less
well understood, but progress is being made. Many recent results based
on analyses of the fundamental plane yield best-fit evolution rates
of $-0.8 < d \log (M/L)_B/dz < -0.5$, and are mutually consistent
within their quoted uncertainties. For example, Treu et al.\ (2001,
2002) find $d \log (M/L)_B/dz = -0.72^{+0.11}_{-0.16}$, van Dokkum et
al.\ (2001) and van Dokkum \& Ellis (2003) find $d \log (M/L)_B / dz =
-0.54 \pm 0.07$, and van der Wel et al.\ (2004) find $d \log
(M/L)_B/dz = -0.71 \pm 0.20$, each based on luminosity-selected
samples out to $z\sim 1$. In addition, Rusin et al.\ (2003; hereafter
R03) analyze a mass-selected sample of gravitational lens galaxies
over the same redshift range and extract a similar evolution rate of
$d \log (M/L)_B / dz = -0.54 \pm 0.09$, a result that is largely
confirmed by van de Ven, van Dokkum \& Franx (2003; hereafter vvF).
The only significant outlier among FP studies is the Gebhardt et al.\
(2003) measurement of $d \log (M/L)_B/dz \simeq -1.0$ to $z=1$ (see
also Im et al.\ 2002). 

Despite the broad consistency of the above evolution rates, there is
some disagreement regarding the implied star formation history of
early-type field galaxies. For example, Treu et al.\ (2001, 2002) and
Gebhardt et al.\ (2003) favor a mean star formation redshift of
$\langle z_f \rangle < 1.5$, which, combined with cluster
results, would appear to confirm the CDM prediction of younger stellar
populations in field E/S0s. The presence of young stars is supported
by spectral (Schade et al.\ 1999; Kuntschner et al.\ 2002; Treu et
al.\ 2002), color (Menanteau, Abraham \& Ellis 2001; vvF) and
evolutionary (van der Wel et al.\ 2004) evidence for some recent star
formation. In contrast, the remaining FP analyses claim a mean star
formation redshift of $\langle z_f \rangle \sim 2$, and little or no
age difference between field and cluster populations. These results
are supported by the spectral study of Bernardi et al.\ (1998) and the
color analysis of Bell et al.\ (2004). Clearly more work is required
to properly interpret current samples of early-type field galaxies,
and account for the possibility that stars form over a range of
redshifts.

The focus of this paper is the gravitational lens sample.  Lenses are
uniquely suited to exploring the luminosity evolution of the
early-type galaxy population in low-density environments (e.g.,
Keeton, Kochanek \& Falco 1998; Kochanek et al.\ 2000; R03). First,
lenses are mass-selected galaxies, and are therefore less susceptible
to Malmquist biases which will favor the inclusion of brighter, bluer
and younger galaxies in luminosity-selected samples. Second, because
the lensing cross section scales strongly with velocity dispersion,
almost all lens galaxies are of early morphological types. Third, the
optical depth is dominated by galaxies in low-density environments,
such as the ``field'' and poor groups (Keeton, Christlein \& Zabludoff
2000). Fourth, lensing naturally selects a galaxy sample out to $z\sim
1$, due to the redshift spread in the source population and competing
volume and cross section effects in the optical depth. Fifth, lenses
constrain projected masses, which can be used to normalize the galaxy
population to a common scale.

Galaxy evolution results based on gravitational lens samples are often
viewed with skepticism, as these investigations (Kochanek et al.\
2000; R03; vvF) do not use measured velocity dispersions in their FP
analyses. There is a legitimate reason for this: with a handful of
exceptions (Foltz et al.\ 1992; Leh\'ar et al.\ 1996; Tonry 1998;
Ohyama et al.\ 2002; Koopmans \& Treu 2002, 2003; Koopmans et al.\
2003; Gebhardt et al.\ 2003; Treu \& Koopmans 2004), the velocity
dispersions of lens galaxies have not been measured. However, the
geometry of the lensed images allows the velocity dispersion to be
estimated very accurately, given a model for the mass distribution. An
isothermal ($\rho \propto r^{-2}$) model is the common choice, as it
is consistent with most constraints on lens galaxies (Kochanek 1995;
Cohn et al.\ 2001; Mu\~noz, Kochanek \& Keeton 2001; Treu \& Koopmans
2002a; Rusin et al.\ 2002; Koopmans \& Treu 2003; Winn, Rusin \&
Kochanek 2003; Rusin, Kochanek \& Keeton 2003; Koopmans et al.\ 2003;
Treu \& Koopmans 2004) and local ellipticals (Fabbiano 1989; Rix et
al.\ 1997; Gerhard et al.\ 2001; see Romanowsky et al.\ 2003 for an
alternative conclusion).  In the context of evolution studies, the
issue is then whether the isothermal assumption is robust and
sufficient.  Criticisms of R03 (Treu \& Koopmans 2002b; Gebhardt et
al.\ 2003) have centered on claims that velocity dispersion
measurements of a few specific lens galaxies require mass profiles
which are different than isothermal.  The implication is that the
results of R03 and similar papers can be discarded on the grounds that
potential systematic and statistical deviations from the isothermal
assumption may significantly affect evolution constraints, but have
not been properly taken into account.  We must therefore determine how
constraints on the luminosity evolution model depend on galaxy
structure, and the observationally permitted range of mass profiles.

The primary goal of this paper is to develop and apply a framework for
simultaneously investigating the structure and evolution of
gravitational lens galaxies. In this way we can consider the range of
mass profiles allowed by the ensemble of lens geometries, and its
effect on stellar population constraints. In essence, we derive limits
on luminosity evolution without any {\em a priori} assumption
regarding the mean mass distribution in early-type galaxies. The
theoretical framework is built upon the self-similar, two-component
mass models introduced and analyzed by Rusin, Kochanek \& Keeton
(2003; hereafter RKK). We show that the formalism is very closely
related to the fundamental plane, and is therefore a powerful tool for
tracing luminosity evolution. The results of our combined
structure/evolution analysis bolster the previous claims of R03 that
older stellar populations are strongly favored for field E/S0s.

A secondary goal of this paper is to better quantify scatter within
the lens galaxy sample, and among early-type field galaxies in
general.  Unlike cluster ellipticals, which typically show little
galaxy-to-galaxy age variance, field ellipticals exhibit significant
scatter about their best-fit ``mean'' evolutionary tracks (e.g., Treu
et al.\ 2001, 2002; RKK; vvF, Treu \& Koopmans 2004; van der Wel et
al.\ 2004), strongly suggesting a range of stellar ages at $z\sim 0$.
Therefore, in addition to investigating single-epoch star formation
models to describe the mean evolutionary history of early-type field
galaxies, we also consider a simple model in which galaxies form over
a range of redshifts. In doing so, we obtain the first stellar
evolution model which reproduces the statistical properties of the
lens sample.

Section 2 outlines the theoretical model for investigating galaxy
structure and evolution. Section 3 details our lens sample and
calculations. Section 4 describes our results. Section 5 compares our
findings with other lensing analyses. In \S 6 we summarize our results
and discuss their implications. We assume $\Omega_M=0.3$,
$\Omega_{\Lambda} = 0.7$ and $H_0 = 65 \kmsm$ for all calculations.

\section{Structure and Evolution in a Common Framework}

\subsection{General Theory}

The geometry of a gravitational lens system yields a precise,
model-independent relationship between the aperture radii defined by the
lensed images and the projected masses they enclose (Schneider, Ehlers
\& Falco 1992). For rings and four-image lenses (quads), the Einstein
radius $R_{Ein}$ is related to the aperture mass $M_{Ein}$:
\begin{equation}
\frac{1}{\pi} \frac{M_{Ein}}{R_{Ein}^2} = \frac{c^2}{4\pi G}
\frac{D_s}{D_d D_{ds}} \equiv \Sigma_{cr} \ .
\end{equation} 
Note that modeling can remove the effects of galaxy ellipticity and
external shear in quads and rings, leaving a negligible ($\sim 2\%$)
uncertainty on eq.~(1). For two-image lenses (doubles), the image
radii $R_1$ and $R_2$ are related to the enclosed masses $M(R_1)$ and
$M(R_2)$:
\begin{equation}
\frac{1}{\pi} \left[ \frac{M(R_1)}{R_1} + \frac{M(R_2)}{R_2}
\right] \left( \frac{1}{R_1 + R_2} \right) =  \Sigma_{cr} \ .
\end{equation}
Modeling cannot robustly remove the effects of galaxy ellipticity and
external shear in doubles. However, we can account for typical
quadrupoles by assuming a $\sim 10\%$ uncertainty on eq.~(2).  The
quantity $\Sigma_{cr}$, defined in eq.~(1), is the critical surface
density, which depends on the angular diameter distance to the lens
($D_d$), to the source ($D_s$), and from the lens to the source
($D_{ds}$). RKK demonstrate that an ensemble of aperture mass-radius
(AMR) relations can be used to statistically constrain the mean mass
profile of lens galaxies, within the context of a model which relates
the mass distribution to the light.

Once a given mass profile has been normalized to satisfy the aperture
mass-radius relation in a particular lens, we can calculate the
line-of-sight stellar velocity dispersion at radius $R$ by solving the
Jeans equation. For constant orbital anisotropy $\beta_{iso} = 1 -
\sigma_{\theta}^2/\sigma_r^2$, the solution is 
\begin{equation}
I(R)\sigma^2(R) = 2 \int_0^{\infty} dz \left(1 -
\beta_{iso}\frac{R^2}{r^2} \right) r^{-2\beta_{iso}} \int_r^{\infty}
du \frac{\nu(u) G M(u)}{u^2} u^{2 \beta_{iso}} \ ,
\end{equation}
where $I(R)$ is the surface brightness and $\nu(r)$ is the
corresponding volume distribution of the luminous matter. The
luminosity-weighted velocity dispersion inside some (circular)
spectroscopic aperture $R$ is then:
\begin{equation}
\langle \sigma^2(R) \rangle = \int_0^R d^2 R^{\prime} I(R^{\prime})
\sigma^2(R^{\prime}) \bigg/ \int_0^R d^2 R^{\prime} I(R^{\prime}) \ .
\end{equation}
Note that galaxy ellipticity and external shear do not significantly
affect the dynamical estimate (Kochanek 1994). The mass normalization
required by the AMR relation makes the stellar velocity dispersion a
very sensitive probe of the radial mass profile in lens galaxies
(e.g., Treu \& Koopmans 2002a; Koopmans \& Treu 2003).

Our ability to estimate the velocity dispersion from the image
geometry means that we can investigate the luminosity evolution of
lenses using the fundamental plane (FP; Djorgovski \& Davis 1987;
Dressler et al.\ 1987):
\begin{equation}
\log R_e = \alpha \log \sigma_c + \beta \langle SB_e \rangle + \gamma
\ ,
\end{equation}
where $R_e$ is the effective (half-light) radius, $\langle SB_e
\rangle$ is the mean absolute surface brightness enclosed by $R_e$,
and $\sigma_c$ is the central stellar velocity
dispersion.\footnote{The quantity $\sigma_c$ is often defined within a
standard aperture of 3\farcs4 (diameter) at the distance of the Coma
cluster (e.g., Jorgensen et al.\ 1996), or some fixed fraction of the
effective radius (typically $R_e/8$).} The FP allows us to predict the
surface brightness that a galaxy would have at $z=0$, given its
velocity dispersion and effective radius. Assuming that the structural
parameters do not evolve (the same assumption which has been made in
all previous evolution analyses of the FP; e.g., Jorgensen et al.\
1996, 1999; Kochanek et al.\ 2000; Treu et al.\ 2001, 2002; van Dokkum
et al.\ 1998, 2001; R03; vvF; Gebhardt et al.\ 2003), the difference
between the observed surface brightness and the predicted $z=0$ value
is then directly related to the luminosity evolution: $\Delta \log
(M/L) = 0.4 \Delta \langle SB_e \rangle = -0.4 \Delta \gamma/
\beta$. This technique -- estimating velocity dispersions from mass
models normalized to match the AMR relations and plugging the values
into the FP -- has been employed by previous studies to trace the
luminosity evolution of lens galaxy samples (Kochanek et al.\ 2000;
R03; vvF). The drawback of these analyses is that they each focussed
on the isothermal mass model, and did not thoroughly consider possible
deviations or the range of permitted profiles.\footnote{Note that
Kochanek et al.\ (2000) experimented with models in which mass traces
light, and found little difference in the favored evolution rate.}

Our goal is to investigate galaxy evolution without any {\em a priori}
assumption regarding the shape of the mass distribution. Of course, we
cannot independently set the profile of each lens galaxy. Instead, we
assume that early-type galaxies form a regular population, in which
the mass distribution is related to the light distribution. We can
then use the FP to simultaneously investigate galaxy structure and
evolution. Specifically, for a given mass profile, the central
velocity dispersion can be calculated for each lens once the profile
normalization has been set by the AMR constraint. These velocity
dispersions can then be compared with the predictions of the
FP/evolution model:
\begin{equation}
\log \sigma_{FP} = \frac{\log R_e - \beta [\langle SB_e \rangle -
2.5 \ e(z)]- \gamma_0}{\alpha} \ ,
\end{equation}
where $\gamma_0$ is the present-day intercept, and $e(z)$ is the
evolution in mass-to-light ratio to the galaxy redshift $z$. For now
we assume that all galaxies are described by a single evolutionary
track; more general models will be considered later.  The
model may then be evaluated by finding the mass profile, evolution
model and zero-point that minimize the scatter of the FP, which is
known to be small at both high and low redshift, and in all
environments (e.g., van Dokkum \& Franx 1996; Jorgensen et al.\ 1996;
Kelson et al.\ 1997; van Dokkum et al.\ 1998, 2001; Pahre, Djorgovski,
\& de Carvalho 1998a; Bernardi et al.\ 2003).

The FP technique described above is very closely related to the
homology formalism of RKK. For a homologous mass distribution, $M =
c_M \, \sigma_c^2 \, R_e / G$, where $c_M$ is a constant. Substituting
the definitions of the surface brightness ($L \propto I_e R_e^2$,
$\langle SB_e \rangle \equiv -2.5 \log I_e$) and implicitly evolving
all luminosities to present-day, we have
\begin{equation}
\log \left(\frac{M}{L}\right) \propto 2 \log \sigma_c + 0.4 \langle
SB_e \rangle - \log R_e + \log c_M\ .
\end{equation}
If we allow the mass-to-light ratio to scale as $L^x = (I_e R_e^2)^x$, then 
\begin{equation}
\log R_e \propto \left(\frac{2}{2x+1}\right)\log \sigma_c + \frac{0.4
(x+1)}{2x+1} \langle SB_e \rangle + \frac{\log c_M}{2x +1 } \ , 
\end{equation}
which looks like the fundamental plane (eq.~5), with $\alpha =
2/(2x+1)$ and $\beta = 0.4(x+1)/(2x+1)$ (see also Faber et al.\ 1987).

The RKK homology formalism (with luminosity evolution) is identical to
the real FP under three reasonable conditions. First, the FP slopes
must be consistent with the definition of $x$, which requires that
$\alpha/\beta = 5/(x+1)$. Note that this condition is met by
measurements of the local FP in the $B$ band, which favor slopes
($1.20 \la \alpha \la 1.25$ and $\beta \simeq 0.32$; Jorgensen et al.\
1996; Bender et al.\ 1998) that are very close to those predicted for
$x=0.3$. Second, we must fit the local intercept $\gamma_0$, which
depends on a combination of $c_M$ and the stellar mass-to-light
ratio. Fitting for $\gamma_0$ is a reasonable procedure, as local
measurements of the FP intercept in low-density environments are
scarce. Third, the velocity dispersion must be measured/estimated in a
spectroscopic aperture which is some fixed fraction of $R_e$. Such a
definition is already used in many FP analyses (e.g., Jorgensen et
al.\ 1996, 1999; van Dokkum et al.\ 1998, 2001; Bernardi et al.\
2003). Note that the size of the spectroscopic aperture relative to
$R_e$ is irrelevant if the intercept is fit, as it changes only the
dynamical constant $c_M$, which has no effect on evolution rates
unless the homology breaks down significantly or evolves with
redshift. The value of $\beta_{iso}$ used to estimate velocity
dispersions is also irrelevant for our analysis, as it can be
incorporated into $c_M$. We make the assumption of constant
$\beta_{iso}$ for computational simplicity. However, the anisotropy is
equally irrelevant for an Osipkov-Merritt (Osipkov 1979; Merritt
1985a,b) parameterization [$\beta_{iso}(r) = r^2/(r^2+r_i^2)$], so
long as we assume that the isotropy radius $r_i$ is proportional to the 
effective radius.

Consequently, the FP and homology frameworks are interchangeable, with
the AMR relations derived from image geometries acting as a suitable
proxy for velocity dispersions. By incorporating stellar evolution
explicitly into the homologous models for galaxy mass distributions,
we can harness the power of the FP to simultaneously constrain the
structure and luminosity evolution of early-type lens galaxies. The
combined analysis allows us to obtain limits on luminosity evolution
that are relatively unbiased with respect to the choice of mass model.
A certain range of mass profiles will be consistent with a tight
fundamental plane, and this range dictates the degree to which a
strong homology might be violated in this galaxy population. The
effect of this range on the evolution estimates can then be taken into
account.

\subsection{Mass Models}

We begin with a two-component model relating the mass and light in
galaxies. The model is parameterized in terms of projected quantities,
as these are directly constrained by gravitational lenses. However,
volume mass distributions are necessary for evaluating the Jeans
equation (eq.~3). We therefore employ three-dimensional analogs
of the models used by RKK.

The mass component which traces the stars is modeled as a Hernquist
(1990) profile, with volume density
\begin{equation}
\rho_{lum}(r) = \frac{\rho_h}{(r/a)(1+r/a)^3} \ ,
\end{equation}
where $a$ is the characteristic radius. In projection, the Hernquist
model closely approximates a de Vaucouleurs surface brightness profile
for $a=0.551 R_e$. The fraction of the total luminosity projected
inside $R$ is denoted $g(R/R_e)$. Allowing the mass-to-light ratio of
the stars to vary with galaxy luminosity ($\Upsilon \propto L^x$), the
projected luminous matter inside an aperture $R$ is then
\begin{equation}
M_{lum}(R) = \Upsilon_{*} L_* \left(\frac{L_{ev}}{L_*}\right)^{1+x}
g\left(\frac{R}{R_e} \right)\ ,
\end{equation}
where quantities are scaled relative to those of an $L_*$ galaxy,
which we define as having a present-day ($z=0$) characteristic
magnitude of $M_*(0) = -19.9 + 5 \log h$ in the rest frame $B$ band
(e.g., Madgwick et al.\ 2002). The quantity $L_{ev}$ is the galaxy
luminosity corrected for stellar evolution, which allows all lenses to
be evaluated at a common age. (see \S 2.3).

The dark matter halo is modeled using the cuspy profile of Mu\~noz et
al.\ (2001), with volume density 
\begin{equation}
\rho_{cdm}(r) = \frac{\rho_c}{ (r/r_b)^n [1+(r/r_b)^2]^{(3-n)/2} } \ ,
\end{equation}
where $r_b$ is the break radius. The profile follows the generalized
form of a simulated CDM halo (Navarro, Frenk \& White 1997; Moore et
al.\ 1999), which has a logarithmic density slope $n$ for $r \ll r_b$
and $3$ for $r \gg r_b$. In most cases we will consider the
``power-law'' limit with $r_b \gg R_e$. The shape of the projected CDM
mass profile is denoted as $m_{cdm}(R/R_e)$, which we normalize such
that $m_{cdm}(2) = 1$. The abundance of CDM is
described by $f_{cdm}$, the fraction of aperture mass in the form of
dark matter inside $R=2R_e$. The total (luminous plus dark) aperture mass
enclosed by $R$ is then
\begin{equation}
M(R) = \Upsilon_*\,L_*\,\left(\frac{L_{ev}}{L_*}\right)^{1+x}\, \left[
g\left( \frac{R}{R_e}\right) + g(2) \frac{f_{cdm}}{1-f_{cdm}}
m_{cdm}\left(\frac{R}{R_e}\right) \right] \ ,
\end{equation}
with $g(2) =0.69$. 

While our mass model is based on the assumption of strong homology,
determining the degree to which early-type galaxies form a homologous
population remains a problem of great interest (e.g., Faber et al.\ 
1987; Caon, Capaccioli \& D'Onofrio 1993; Bertin et al.\ 1994, 2000;
van Albada, Bertin, \& Stiavelli 1995; Ciotti, Lanzoni \& Renzini
1996; Pahre, de Carvalho \& Djorgovski 1998b; Gerhard et al.\ 2001;
Borriello, Salucci \& Danese 2003; Trujillo, Burkert \& Bell 2004).
For example, there is strong evidence that the total mass-to-light
ratio increases with luminosity (e.g., Gerhard et al.\ 2001; Bernardi
et al.\ 2003; Borriello et al.\ 2003; RKK; Padmanabhan et al.\ 2004),
but little consensus as to the source of this effect.  Gerhard et al.\ 
(2001) argue that we are seeing a luminosity dependence of the stellar
mass-to-light ratio.  In contrast, Padmanabhan et al.\ (2004) claim
that the effect is due to an increasing dark matter abundance with
galaxy luminosity, while the stellar populations exhibit negligible
variation (see also Kauffmann et al.\ 2003). We have tested a
non-homologous mass distribution, in which the dark matter abundance
$f_{cdm}$ varies with galaxy luminosity as
\begin{equation} \frac{f_{cdm}}{1-f_{cdm}} \equiv
  r_{cdm} = r_{cdm,*}\left( \frac{L_{ev}}{L_*} \right)^y \ , 
\end{equation}
where $r_{cdm} \equiv M_{cdm}(2R_e)/M_{lum}(2R_e)$. Unfortunately, the
tests show that our lensing analysis has no power to distinguish even
the extreme cases of $y=0$ (structurally homologous) and $x=0$
(systematically non-homologous), as we can measure only a combination
of parameters $x + y$.  Specifically, we obtain a virtually identical
value for $x$ if $y=0$ as we do for $y$ if $x=0$, with no difference
in the goodness of fit or the favored evolution model. This result is
primarily due to the fact that lenses measure total masses only.
Consequently, homologous mass distributions contain all the
phenomenology we can probe: namely, the concentration of the mass
distribution and the dependence of the total mass-to-light ratio on
luminosity (rather than the individual behavior of either mass
component). The evolution estimates are the same under either
structural assumption.

Strict homology is also relaxed if galaxies exhibit structural scatter
at fixed luminosity. This could be tested by modeling the galaxy
population with a statistical distribution of mass profiles, much as
we will do for star formation redshifts (see \S 2.3). In practice,
however, this is difficult to accomplish within the context of our
global mass model. For a given galaxy, different mass profiles require
different stellar mass-to-light ratios, which renders impossible a
parameterization of the galaxy population in terms of a single value
of $\Upsilon_*$. We choose to forgo the direct modeling of profile
scatter, as we do not wish to add more nuisance parameters to cloud
the analysis. Moreover, in \S 4.2 we show that structural scatter is
not necessary to produce a good statistical description of the
data. Hence, a single profile model satisfies Occam's razor. By
ignoring structural scatter, we will maximize the range of formation
redshifts necessary to account for the scatter observed about the AMR
relations.

\subsection{Evolution Models}

We account for evolutionary effects by converting each observed galaxy
luminosity $L(z)$ to a common stellar age, yielding an
evolution-adjusted luminosity $L_{ev}$. This may be accomplished in two
ways, depending on the complexity of the star formation model.

To derive constraints on the mean star formation redshift and
evolution rate, we assume that the galaxies are described by a single
star formation redshift (Model 1). For such a model we can normalize
the galaxy luminosities at any redshift we desire, and simplicity
dictates that we evolve them to $z=0$:
\begin{equation} 
\log L_{ev} = \log L(0) = \log L(z) + e(z) ,
\end{equation} 
where $e(z)$ is the change in mass-to-light ratio out to redshift $z$.
We describe luminosity evolution with a pair of models.  First, there
is a linear model in which $e(z) = [d \log (M/L)_B /dz] \ z$. This is a
useful approximation for older stellar populations, and offers a
standard for comparing results from different analyses.  Second,
$e(z)$ can be set according to the detailed evolutionary tracks of
stellar population models, which we calculate using the GISSEL96
version of the Bruzual \& Charlot (1993) synthesis code (hereafter
denoted ``BC96''). This technique facilitates constraints on the mean
star formation redshift.  A $\chi^2$ statistic is sufficient to
evaluate Model 1.

Modeling the early-type galaxy population with a single formation
redshift or evolution rate rarely provides a good statistical
description of the data (e.g., vvF; van der Wel et al.\ 2004).
Therefore, we also investigate a model in which star formation takes
place over a range of redshifts (Model 2). This is the first time that
such a model has been applied to a gravitational lens sample. As in
Model 1, galaxies are formed obeying the self-similarity relation
(eq.~12), though in this case they will exhibit a range of ages at any
given redshift.  Because we want to evaluate the galaxies at a fixed
age, we use a stellar evolution model to convert the observed
luminosity to its value at formation: $L_{ev} = L(z_f$). Note that in
this context $\Upsilon_*$ represents the characteristic mass-to-light
ratio at $z_f$. We consider a model in which star formation takes
place between $z_{f,min}$ and $z_{f,max}$, with uniform probability
density in $\log z_f$: $dP/d\log z_f = {\rm constant.}$ The added
complexity of Model 2 requires a likelihood formalism for analysis.

\section{Data and Analysis}

\subsection{Lens Sample and Parameters}

We reanalyze the gravitational lens sample of RKK, which includes 22
early-type galaxies and bulges with measured redshifts. The raw
photometric (total magnitudes and effective radii from Hubble Space
Telescope observations) and geometric (Einstein radii for ring and
quad lenses; image radii for doubles) parameters are listed in R03 and
RKK, respectively. Assuming an isothermal profile, our lens galaxy sample
is characterized by a mean velocity dispersion of $\sim 240 \kms$. 

For each galaxy, we convert the observed magnitudes ($m_{obs,Y}$ in
filter $Y$) into a rest frame $B$-band magnitude using the procedure
outlined in R03. This involves using the BC96 spectral evolution code
to calculate a spectral energy distribution (SED) as a function of
redshift, which is then convolved with transmission curves for HST
filters (Holtzman et al.\ 1995). We construct synthetic ``colors''
$C(B,Y) \equiv m_{mod,B} - m_{mod,Y}$ between rest frame magnitudes in
filter $B$ and directly measurable magnitudes in filter $Y$ for a
galaxy at the appropriate redshift. The rest frame magnitude $m_B$ is
then
\begin{equation}
m_{B} = \frac{\sum_Y[m_{obs,Y}+C(B,Y)]/(\delta m_{obs,Y})^2}{\sum_Y
1/(\delta m_{obs,Y})^2} \ ,
\end{equation}
where the sum is taken over all filters in which the galaxy has been
observed. The model SED depends on the star formation history,
metallicity, initial mass function (IMF) and cosmology. We take as a
fiducial model an instantaneous star burst at $z_f=3$ with solar
metallicity $Z=Z_{\odot}$ and a Salpeter (1955) IMF, along with our
standard cosmological parameters. For a given spectrophotometric
model, errors on the derived rest frame magnitudes tend to be small
($\la 0.1$ mag), as most lens galaxies in our analysis sample have
good photometry in at least one filter (see R03). However, the
conversion does depend on the assumed spectral template. We
account for this in our error budget by considering the scatter
induced by a broad ensemble of plausible models ($1 < z_f < 5$, $0.4
Z_{\odot} < Z < 2.5 Z_{\odot}$). This results in an additional
uncertainty of $0.1 - 0.15$ mag in the interpolated magnitudes.
Following R03 and RKK, we therefore set a uniform error of $0.2$ mag
on the derived rest frame magnitudes, or $\delta \log L = 0.08$. We
will model these and all subsequent errors as Gaussian.

It is important to note that the assumed value of the luminosity error
has a limited effect on our analysis. For Model 1, our determination
of whether the model is a good statistical description of the data is
based on the best-fit $\chi^2$, which of course depends on $\delta
\log L$. Because a model with a single star formation redshift is
expected to perform poorly, we obtain constraints on individual
parameters (using standard $\Delta \chi^2$ limits) by rescaling all
errors such that the best-fit $\chi^2$ is equal to the number of
degrees of freedom. Post-rescaling, the initial value of $\delta \log
L$ is of little importance. The rescaling procedure yields constraints
similar to those derived from a jackknife error analysis or bootstrap
resampling (each without error rescaling), as all three techniques are
designed to provide error bars which are consistent with the observed
scatter. For simplicity we will quote only the $\Delta \chi^2$
results. In Model 2 we seek a good fit to the data in the absolute
sense (see \S 3.2), so error rescaling will not be used.  While our
parameter constraints do vary with $\delta \log L$, our ability to
derive a statistically consistent model does not.  We have tested
several values ($0.02$, $0.04$, $0.08$) of $\delta \log L$, and find
that each yields models which provide a good statistical description
of the data.

More important than the assumed value of $\delta \log L$ is that we
apply it uniformly to all lenses. This ensures that each galaxy is an
important contributor to the fit, which allows us to better
constrain the ``mean'' properties of the population. As discussed in
\S 5.1, uniform luminosity errors should reduce biases in
constraining evolutionary models. 

The errors in the radii entering the AMR relations are negligible, and
we can safely ignore them. Uncertainties in the model-predicted masses
or velocity dispersions are dominated by the luminosity errors, which
are summarized above. Parameters correlated with the effective radius
are treated as in RKK, though these effects are also negligible.  Four
of the systems in our sample do not have measured source redshifts.
For these we assume $z_s = 2.0\pm 1.0$ and derive uncertainties in
$\log \Sigma_{cr}$ according to the Monte Carlo procedures detailed in
RKK.  Finally, because unconstrained quadrupoles effectively smear the
AMR relation for doubles, we include an additional tolerance of 10\%
in mass (or 5\% in velocity dispersion) for such cases.

\subsection{Calculations}

The AMR relations provide constraints on the homologous mass model, or
equivalently, on the FP. For this purpose we introduce a function to
describe the offset of each galaxy from its relevant AMR relation
(eq.~1 for quads and rings, eq.~2 for doubles): 
\begin{eqnarray}
\Delta_{AMR,i} \equiv 
\left\{
\begin{array}{ll}
\log M_{mod}(R_{Ein,i}) - \log (\Sigma_{cr,i} \pi R_{Ein,i}^2 ) & {\rm \, \, rings, quads}\\
\log (M_{mod}(R_{1,i})/R_{1,i} +
M_{mod}(R_{2,i})/R_{2,i}) - \log(\Sigma_{cr,i} \pi (R_{1,i} +
R_{2,i})) & {\rm \, \, doubles}
\end{array}  
\right. 
\end{eqnarray}

We constrain Model 1, which postulates a single mass profile and a
single star formation redshift for the lens galaxy population, by
optimizing the goodness-of-fit parameter
\begin{equation} 
\chi_{AMR}^2 \equiv \sum_i \chi^2_{AMR,i} =   \sum_i \left( \frac{\Delta_{AMR,i}}{\delta_{scl} \delta_{AMR,i}}  \right)^2 \ , 
\end{equation} 
where $\delta_{scl}$ is the error rescaling factor. The logarithmic
uncertainty $\delta_{AMR,i}$ on each data point is derived using the
Monte Carlo methods outlined in \S 3.1, but is well approximated as
$\delta_{AMR,i}^2 \simeq (1+x)^2 (\delta \log L)^2 + (\delta \log
\Sigma_{cr,i})^2 + \delta^2_{\gamma}$, where $\delta_{\gamma}$ is the
additional 10\% tolerance for quadrupole-related smearing of the
mass-radius relation in doubles. We set $\delta_{\gamma} = 0$ for
quads and rings, and $\delta \log \Sigma_{cr,i} = 0$ for systems with
a measured $z_s$.

Alternatively, we could use the FP approach and evaluate Model 1 using the
goodness-of-fit function
\begin{equation}
\chi^2_{FP} = \sum_i \left[ \frac{\log \sigma_{FP,i} - \log
\sigma_{c,i}}{\delta_{scl} \delta_{FP,i}} \right]^2 \ ,
\end{equation}
where $\sigma_{c,i}$ is the velocity dispersion calculated from the
AMR-normalized mass profile, and $\sigma_{FP,i}$ is the value
predicted from the FP/evolution model (eq.~6). The uncertainty
parameter is $\delta_{FP,i} = 0.5 \, \delta_{AMR,i}$. While this looks
like a very different formalism, it is not. Analyzing estimated
velocity dispersions with the FP (eq.~18) yields numerically identical
results to analyzing AMR relations with the homology model (eq.~17).
Only if the local intercept $\gamma_0$ is known to high systematic and
statistical accuracy ($\ll 0.1$) does using the FP formalism add any
information. However, to make use of the local zero-point, one must
also know the slopes of the FP and the orbital anisotropy. In
addition, the velocity dispersion must be estimated in a spectroscopic
aperture identical to the one used to derive the local FP parameters.
We therefore find that the mass homology relation for lenses contains
all the information of the FP with fewer systematic errors given the
direct relationship of lens data to mass rather than velocity
dispersion.

Following optimization of the parameters, we set $\delta_{scl}$ so
that the best-fit model has $\chi^2 = N_{DOF}$, the number of degrees
of freedom. This preserves the relative weighting among the data
points, which naturally gives less weight to doubles and those systems
with an estimated $z_s$, and therefore does not alter the optimized
values of the model parameters. However, rescaling does allow us to
relate the uncertainties to the observed scatter in the model,
yielding parameter errors which agree with those derived from the
bootstrap and jackknife techniques (see \S 3.1). Rescaling is
particularly important because Model 1 is undoubtedly a simplistic
representation of galaxy structure and evolution. Hence, the $\chi^2$
is likely to be dominated by unmodeled complexity (i.e., deviations
from self-similarity and an intrinsic spread in star formation
redshifts) in the galaxy population, rather than by observational
errors.

Model 2 postulates a homologous mass profile for the lens sample, but
allows for star formation over a range of redshifts. This adds significant
complexity to the analysis. Specifically, while a single formation
redshift yields a single value of $\Delta_{AMR}$ for each galaxy (given a
fixed set of structural parameters), a distribution of formation redshifts
yields a distribution $dP/d\Delta_{AMR}$. Because a $\chi^2$ statistic
alone is insufficient for evaluating Model 2, we turn to a likelihood
formalism. We evaluate the likelihood $\ln \scriptL = \sum_i \ln
\scriptL_i$, where 

\begin{equation} 
\scriptL_i =  \int_{z_{f,0}}^{z_{f,max}} \exp 
\left(-\frac{\chi_{AMR,i}^2}{2} \right)  \frac{dP}{d z_f} d z_f \bigg/
 \int_{z_{f,0}}^{z_{f,max}}  \frac{dP}{d z_f} d z_f \ , 
\end{equation} 
and $\chi_{AMR,i}^2$ is defined in eq.~(17). The lower limit of
integration is $z_{f,0} = {\rm max} (z_{f,min}, z_i)$, where $z_i$ is
the redshift of the $i$th lens galaxy. 
Confidence limits on parameters are derived in the
usual manner for maximum likelihood methods by using differences in $\ln
\scriptL$.  There is no rescaling of errors in Model 2. Note that the
likelihood formalism (eq.~19) should reduce to the $\chi^2$ formalism
(eq.~17) in the limit of a single star formation redshift (Model 1).  We
have confirmed that the techniques yield identical parameter constraints
in this case, so long as we keep $\delta_{scl} = 1$.

In addition to finding the best-fit model, we also want to determine
whether it is a good fit to the lens data in the absolute sense. For
Model 1, we simply compare the best-fit $\chi^2$ (for
$\delta_{scl}=1$) to the number of degrees of freedom. Model 2,
however, requires a more intricate, Monte Carlo-based approach in
which we select and analyze simulated lens samples, which are based on
the geometries of our actual lenses. First, we randomly select the
formation redshift for each galaxy according to the statistical
distribution $dP/d z_f$, and set the formation luminosity such that
the homologous mass model perfectly reproduces the AMR relation. The
luminosity is then evolved to the measured galaxy redshift, and
perturbed according to the assumed luminosity errors.  Uncertainty in
the mass scale due to an estimated source redshift is handled by an
additional perturbation.  We calculate the likelihood of each simulated
data set given the model ($\ln \scriptL_{sim}$), and build up an
expected likelihood distribution ($dP/d \ln \scriptL_{sim}$), which
we crudely parameterize by its mean $\langle \ln \scriptL_{sim}
\rangle$ and standard deviation $\ln \scriptL_{sim}^{STD}$. If the
real lens sample is well described by the model, we would expect its
likelihood ($\ln \scriptL$) to be consistent with the distribution
$dP/d \ln \scriptL_{sim}$. We define consistency as $|\ln \scriptL -
\langle \ln \scriptL_{sim}\rangle| \leq \ln \scriptL_{sim}^{STD}$.
Finally, we note that these Monte Carlos test a necessary -- but not a
sufficient -- condition for compatibility between samples (in this
case, real and simulated data).

\section{Results}

\subsection{Focus on Structure}

We begin by updating the RKK analysis of galaxy structure, taking into
account the effects of galaxy evolution. To facilitate comparisons
with the RKK results, we will derive our structural constraints using
Model 1, which postulates a single star formation redshift. We assume
the linear evolution model, but replacing this with the detailed
evolutionary tracks has little quantitative effect. In Fig.~1 we plot
constraints on the dark matter abundance $f_{cdm}$ and inner power-law
slope $n$ by simultaneously optimizing the other parameters -- $x$, $d
\log (M/L)_B/dz$ and the normalization ($\gamma_0$ for the FP,
$\Upsilon_*$ for the homology formalism). Solid lines are the 68\%
($\Delta \chi^2 < 2.30$) and 95\% ($\Delta \chi^2 < 6.17$) confidence
limits in two dimensions; dotted lines are the 68\% ($\Delta \chi^2 <
1$) and 95\% ($\Delta \chi^2 < 4$) confidence limits in one dimension.
These and all subsequent values of $\Delta \chi^2$ follow the setting
of the scale factor $\delta_{scl}$ such that the best-fit model has
$\chi^2 = N_{DOF}$. Constraints are shown for three different values
of the CDM break radius, $R_b/R_e = 3$, $10$ and $50$.  Smaller values
of $R_b/R_e$ mean that the steep outer slope has a more prominent
effect on the radial scales probed by the lens sample ($0.2 < R/R_e <
7$; see RKK). To compensate, the inner slope $n$ must become shallower
as $R_b / R_e$ is decreased. The best-fit models have $\chi^2 = 44.1$,
$43.8$ and $44.0$ for $R_b/R_e = 3$, $10$ and $50$, respectively,
corresponding to an rms scatter of $0.16$ in $\log (M/L)$. Based on
our estimated uncertainties, this implies an intrinsic scatter of
$\sim 30\%$. Note that a $\chi^2$ of $44$ for $N_{DOF} = 17$ is a very
poor fit, suggesting that structural homology and a single star
formation redshift do not provide a good description of the lens data
(see also vvF). This is also the case for the standard FP, whose
intrinsic scatter is larger than can be explained by measurement
errors alone (e.g., Jorgensen et al.\ 1996).

Our nearly identical values of $\chi^2$ for the different break radii
indicate that our model is over-parameterized. For simplicity we will
set $R_b/R_e = 50$ for all subsequent calculations, and therefore
treat the dark matter halo as a power law. Note, however, that this
limit is not identical to the power-law surface mass $M(R)\propto
R^{3-n}$ which RKK used to describe the halo.  Specifically, the RKK
model does not have a consistent three-dimensional projection for
$n\leq 1$. As $n\rightarrow1$ the two models are quantitatively
different, with the RKK power law approaching a mass sheet ($\Sigma =
{\rm constant}$), and the cuspy model approaching a logarithmic surface
density distribution ($\Sigma \propto \ln R$). Hence, for a given
point ($f_{cdm}$ and $n\simeq 1$) in the parameter plane describing
our new model, the projected mass profile is slightly steeper than the
same point in the RKK model plane.

Despite our inability to separate the effects of the inner power-law
slope, break radius and CDM mass fraction, all acceptable models are
very close to isothermal on the scale of several effective radii. We
plot the mass profiles in Fig.~1. Our profile constraints agree with
those of RKK, who assumed a fixed evolution model corresponding to a
star formation redshift of $z_f=3$. By including luminosity evolution
as a free parameter, however, we allow for a slightly broader range of
mass profiles. In the limit of a nearly scale-free mass model
($\Upsilon_* \rightarrow 0$, $f_{cdm} \rightarrow 1$, $R_b/R_e \gg
1$), our constraint on the logarithmic density slope is $n=2.06 \pm
0.17$, compared to the RKK limit of $n=2.07\pm 0.13$.\footnote{Using
  Model 2 we obtain $n=2.05 \pm 0.11$, which is consistent with the
  results from Model 1. The absence of error rescaling in Model 2 (see
  \S 3.2) accounts for most of the difference in the profile
  uncertainty. Consequently, we conclude that allowing for a range of
  star formation redshifts has little effect on the mean mass profile
  favored by the lens sample.}  Each constraint is at 68\% confidence,
or $\Delta \chi^2 < 1$. We still detect dark matter to very high
significance, as a model in which mass traces light is rejected at
$\Delta \chi^2 = 9.2$ (compared to $\Delta \chi^2 = 10.7$ in RKK). In
summary, uncertainties in the luminosity evolution do not affect our
conclusions that early-type galaxies are nearly isothermal on the
scale of a few effective radii, and that a small but non-zero fraction
of dark matter ($f_{cdm} > 0.08$, 95\% C.L.) is required in this
regime, independent of the stellar mass-to-light ratio.

The zero-point of the FP depends on some combination of the radial
mass profile, mass-to-light ratio, orbital anisotropy and
spectroscopic aperture, as these factors determine the dynamical
constant $c_M$ (see eq.~7 and 8). If the local zero-point $\gamma_0$
has been accurately measured for a given set of slopes $\alpha$ and
$\beta$, then the FP formalism adds information that is not included
in the homology formalism, and could therefore improve our model
constraints. We refit the intercept in our analysis because there is
no accurate, independent measurement of $\gamma_0$ specifically for
low-density environments. However, we can estimate the precision to
which $\gamma_0$ must be measured in order to affect constraints on
our structure/evolution model. In Fig.~2 we show the best-fit value of
$\gamma_0$ as a function of mass profile. For demonstration purposes
we assume FP slopes corresponding to $x=0.3$ ($\alpha = 1.25$ and
$\beta = 0.325$), and calculate velocity dispersions for isotropic
orbits ($\beta_{iso} = 0$) in an aperture with radius $R_e / 8$. We
find that $\gamma_0$ varies slowly over the parameter plane if the FP
slopes are fixed. The values vary wildly if $x$ is simultaneously fit.
The value of $\gamma_0$ must be measured to a precision of $\ll 0.1$
to improve the model constraints. We therefore find that the FP adds
little useful information beyond that already contained in the AMR
relations.

While we have assumed that early-type galaxies are homologous,
constraints on the typical mass profile (Fig.~1) show the range over
which individual galaxy profiles might reasonably scatter.  Dynamical
studies of individual lenses can directly address departures from
homology, and are therefore a powerful complement to our statistical
analysis. Velocity dispersions have been measured for seven of the
lenses in our sample: Q2237+0305 (Foltz et al.\ 1992), MG1549+3047
(Lehar et al.\ 1996), PG1115+080 (Tonry 1998), HST14176+5226 (Ohyama
et al.\ 2002; Gebhardt et al.\ 2003; Treu \& Koopmans 2004),
MG2016+112 (Koopmans \& Treu 2002), 0047--281 (Koopmans \& Treu 2003),
and B1608+656 (Koopmans et al.\ 2003). Two of the galaxies are
particularly interesting, as dynamical measurements would imply that
they have mass profiles which deviate significantly from isothermal.
PG1115+080 (Weymann et al.\ 1980) has a velocity dispersion which is
higher than the isothermal prediction, indicating a steeper mass
profile (Tonry 1998; Treu \& Koopmans 2002b). In contrast,
HST14176+5226 (Ratnatunga et al.\ 1995) has a velocity dispersion
which is lower than the isothermal prediction (Ohyama et al.\ 2002;
Gebhardt et al.\ 2003; Treu \& Koopmans 2004), indicating a shallower
mass profile. Based on the existence of such structural outliers, it
has been argued that estimated velocity dispersions are insufficient
for evolution studies with gravitational lens samples. We now show
that this will not be a significant concern for our new evolution
constraints (\S 4.2).

We wish to determine the degree to which dynamical measurements of
lens galaxies are compatible with the range of mass profiles allowed
by our FP/homology analysis. First, at each point in $(f_{cdm}-n)$
space, we set the profile normalization of the lens according to its
AMR relation. We then estimate the velocity dispersion and compare it
to the measured value. For our dynamical estimates we use a circular
aperture with an area equal to that of the aperture in which the
velocity dispersion has been measured, and assume isotropic orbits
($\beta_{iso} = 0$). We expect our comparisons to be accurate at the
level of $\sim 15 \kms$. For each lens, we plot in Fig.~3 contours of
constant $\sigma$, along with our global profile constraints.  We see
that five of the seven lenses are nearly isothermal, and are
consistent with our 68\% confidence limits on the mean mass profile.
Note that while our constraints on the individual mass profiles are
qualitatively similar to those derived in the literature, there are
small systematic disparities due to the use of differing structural
parameters (effective radius, image separation, or environmental
corrections). We confirm that the dynamical measurements of two
galaxies require profiles that are formally inconsistent with
isothermal. However, the comparison ignores systematic uncertainties
in the dynamical model due to the derivation of velocity dispersions
from Gaussian profiles and the ellipticities of the galaxies, each of
which will reduce the significance of these deviations (see Kochanek,
Schneider \& Wambsganss 2004 for additional discussion). For
PG1115+080, Tonry (1998) measures $\sigma = 281 \pm 25 \kms$ in a
$1\farcs0$ squared aperture. We find that this galaxy must have a
profile which is significantly steeper than isothermal: $\rho \propto
r^{-2.4}$, similar to the value favored by Treu \& Koopmans
(2002b).\footnote{Note that we have a much smaller effective radius
  ($r_e \simeq 0\farcs5$ versus $0\farcs8$), and this accounts for our
  acceptance of the constant mass-to-light ratio model which Treu \&
  Koopmans (2002b) nominally rejects.} For HST14176+5226, Gebhardt et
al.\ (2003) measure $\sigma = 202 \pm 9 \kms$ in a similar aperture,
which is marginally inconsistent with a previous measurement of
$\sigma = 230 \pm 14 \kms$ by Ohyama et al.\ (2002) and a similar
measurement by Treu \& Koopmans (2004). We plot the Gebhardt et al.\ 
(2003) value because it implies the greatest departure from the mean
model. As expected, we find that HST14176+5226 requires a profile that
is shallower than isothermal, with $\rho \propto r^{-1.7}$, and a
rather high dark matter fraction ($f_{cdm} > 0.6$). The discrepancy
persists if we substitute the Ohyama et al.\ (2002) or Treu \&
Koopmans (2004) measurements, though at a reduced significance.  While
both PG1115+080 and HST14176+5226 appear to have mass profiles which
are different than isothermal, Fig.~3 shows that the constraints on
their profiles are consistent with our 95\% confidence limit on the
lens galaxy population -- though only marginally consistent with our
68\% confidence range. However, if the width of our statistical
constraints reflects intrinsic scatter about the mean homology
relation, then we might expect $\sim 1/3$ of lenses to be outliers at
the level of PG1115+080 or HST14176+5226.  This fraction is consistent
with the present dynamical sample.

Consequently, our profile constraints are a
sufficient proxy for measured velocity dispersions in gravitational
lens galaxies. By simultaneously considering structure and evolution,
we will derive limits on the evolution rate and mean star formation
redshift which account for the allowed range over which galaxies may
depart from a strict homology.

\subsection{Focus on Luminosity Evolution}

We now turn to luminosity evolution. Because of the large degeneracy
between the CDM slope $n$ and abundance $f_{cdm}$, we will further
simplify the total mass model to the scale-free limit ($\Upsilon_*
\rightarrow 0$, $f_{cdm} \rightarrow 1$, $R_b/R_e \gg 1$).  Note that
this simplification does not have any effect on the evolution
constraints versus the full mass model, but it does make it easier to
display the results, particularly with regard to the interplay between
mass concentration and evolution.

\subsubsection{Model 1: Single Star Formation Redshift}

We begin by analyzing Model 1, which postulates that lens galaxies
were formed at a common epoch, and therefore follow a common
evolutionary track as a function of redshift. This model is intended
to describe the mean properties of the galaxy sample, and provide a
standard for comparing results obtained by different groups.  We
reiterate, however, that this model does not provide a statistically
good fit to the data. As we show in \S 4.2.2, including a range of
star formation redshifts leads to a vastly improved model of the
lens sample (see also vvF).

In Fig.~4 we plot pairwise constraints on the evolution rate $d \log
(M/L)_B/dz$, mass-to-light ratio index $x$, and logarithmic density
slope $n$, which now describes the total mass distribution. As we can
see from Fig.~4, the parameters in our model are correlated. The
anti-correlation between $x$ and $d \log (M/L)_B/dz$ is easiest to
explain, as these two quantities enter the model as a product: $\log
M(R) \propto (1+x)[\log L(z) + d \log (M/L)_B/dz \ z]$ (see eq.~12).
This effect widens the favored range of $x$, compared to the
constraints derived by RKK for a fixed $z_f = 3$ evolution model.
Specifically, we find $x=0.18^{+0.24}_{-0.18}$, while the RKK value is
$x=0.14^{+0.16}_{-0.12}$ (both 68\% C.L.). The source of the other
correlations is not clear, though they may be related to the
increasing mass scale of the lens sample with redshift (R03).

Optimizing over all structural parameters, we measure an evolution
rate of $d \log (M/L)_B/dz = -0.50\pm 0.19$ at 68\% confidence.  The
95\% confidence range is $-0.86 < d \log (M/L)_B/dz < -0.10$. These
constraints are identical for the full mass model and the scale-free
limit, and are not significantly altered by setting $x$ to reflect
locally measured FP slopes. Our results are consistent with the
findings of R03, who measured $d \log(M/L)_B/dz = -0.54 \pm 0.09$
(68\% C.L.)  from a sample of 28 lens galaxies under the assumption of
isothermal mass profiles. Simultaneously optimizing the galaxy
structure does weaken constraints on the evolution rate, but not to
the point that they are uninteresting. The conclusion that lenses
favor a slow evolution rate is unchanged.

We next consider bounds on the mean star formation redshift by using
evolution tracks derived from the BC96 population synthesis
models. Pairwise constraints are plotted in Fig.~5. The constraint on
the mass-to-light ratio index ($x = 0.14^{+0.22}_{-0.16}$, 68\% C.L.)
is somewhat tighter than that derived using the linear evolution rate.
The small difference is due to the fact that the real tracks do not
map into the linear tracks for later formation redshifts; e.g., there
is no real track which is well approximated by a linear model for $z_f
\la 1.5$.

The best-fit model has a mean star formation redshift of $\langle z_f
\rangle \simeq 2.2$, and the data are consistent with stellar ages as
old as the universe itself. Coming from the other side, the
constraints are far more restrictive: $\langle z_f \rangle > 1.8$ at
68\% confidence, and $\langle z_f \rangle > 1.5$ at 95\%
confidence. Note that these results are not sensitive to our
photometric conversion technique (\S 3.1). We find quantitatively
similar bounds if the luminosities are derived from a $z_f = 1.5$
spectrum, or from the spectra which best fit the colors of individual
lenses. Excluding the four systems with estimated source redshifts
slightly weakens the 95\% confidence limit to $\langle z_f \rangle >
1.4$.

The above evolution constraints are remarkable, considering that we
have assumed no specific shape for the mass profile of lens galaxies.
While we do assume that early-type galaxies comprise a regular
population, the range of permitted profiles crudely approximates the
degree of scatter allowed by the AMR relations. We can therefore
account for possible departures from homology, without directly
measuring individual velocity dispersions. Based on this simultaneous
analysis of structure and evolution, we conclude that, on average,
early-type galaxies in low-density environments have rather old
stellar populations ($\langle z_f \rangle \simeq 2$), similar in age
to their cluster counterparts. 

\subsubsection{Model 2: Range of Star Formation Redshifts}

We now turn to Model 2, which allows galaxies to form over a range of
redshifts: $dP/d\log z_f = {\rm constant}$ between $z_{f,min}$ and
$z_{f,max}$. (Substituting a model in which the probability density of
star formation is uniform over $z_f$, rather than $\log z_f$, has
little effect on any of our primary results.) In Fig.~6 we plot
constraints on the star formation model, optimizing over the global
mass profile and mass-to-light ratio parameters. We have re-mapped the
parameter space in terms of $(z_{f,min} + z_{f,max})/2$ and
$(z_{f,max} - z_{f,min})/2$, as this better illustrates the physical
meaning of the constraints. In the main panel (upper left), solid
lines are the 68\% ($\Delta \ln \scriptL = -1.15$) and 95\% ($\Delta
\ln \scriptL = -3.09$) confidence limits in two dimensions; dotted
lines are the 68\% ($\Delta \ln \scriptL = -0.5$) and 95\% ($\Delta
\ln \scriptL = -2.0$) confidence limits in one dimension. In
subsequent panels of Fig.~6 we overlay contours to better describe the
distribution of star formation redshifts in the model. Specifically,
we plot the minimum, maximum and mean formation redshifts, as well as
the fraction of stars forming earlier than $z_f = 2.0$ and later than
$z_f = 1.5$.

The lens sample favors models with significant scatter in star
formation redshift: $z_{f,max}-z_{f,min} > 0.4$ at 95\% confidence and
$z_{f,max}-z_{f,min} > 1.7$ at 68\% confidence, each based on
one-dimensional likelihood differences. The required scatter in $z_f$
increases with mean star formation redshift. This can be understood by
noting that luminosity evolution depends much more sensitivity on
$z_f$ when $z_f$ is small. At higher mean formation redshifts, a
broader redshift range is required to produce the same luminosity
scatter for galaxies at $z<1$. The nearly linear nature of the
degeneracy -- $(z_{f,max}-z_{f,min})/2 \propto
(z_{f,min}+z_{f,max})/2$ -- indicates that the analysis is most
sensitive to $z_{f,min}$, which is consistent with the rapid variation
of evolutionary tracks for late formation redshifts. For the lens
sample, $z_{f,min}$ cannot be much smaller than 1. Undoubtedly the
high redshift ($z>0.5$) lens galaxies will dominate this constraint,
as they are most sensitive to $z_f$. In contrast, models with much
later star formation could be allowed if only low redshift lenses were
considered. We note that the distribution of star formation redshifts
can vary significantly among acceptable models of the lens sample. For
example, the most favored models ($\Delta \scriptL < -1.15$) allow
anywhere from $0$ to $40\%$ of galaxies to form at $z_f< 1.5$
(Fig.~6). We therefore conclude that low-redshift star is an important
component of our model.  However, in all of our acceptable models the
mean star formation redshift is relatively high ($\langle z_f \rangle
> 1.5$). 

It is interesting to note that allowing a range of star formation
redshifts leads to a relatively modest improvement versus a
single-$z_f$ scenario, at least from a maximum likelihood perspective.
For example, we find that the best-fit single-$z_f$ model ($z_{f,min}
= z_{f,max} \sim 2.2$) is at only $\Delta \ln \scriptL = -2.1$ with
respect to the best overall model. There is, however, a large
improvement in terms of finding models which provide a good
statistical description of the data. As the spread in $z_f$ is
increased from zero, the likelihood describing the data ($\ln
\scriptL$) increases slowly, while the mean likelihood of data sets
drawn from the Monte Carlo ($\langle \ln \scriptL_{sim} \rangle$)
decreases due to the fact that the model is inherently more diffuse.
This leads to a rapid improvement in matching the statistical
properties of the lens sample to that of the models. For example, the
best-fit single-$z_f$ model has a likelihood which is lower than its
corresponding $\langle \ln \scriptL_{sim} \rangle$ by $| \Delta \ln
\scriptL | \sim 12$ (this is nearly equal to $(\chi^2 - N_{DOF})/2$, as
expected), and well outside the range described by $\ln
\scriptL_{sim}^{STD}$. Models with a single star formation redshift
are thus clearly inconsistent with the statistical properties of the
data.  However, by $z_{f,max}-z_{f,min} \sim 1$ we find models for
which $|\ln \scriptL - \langle \ln \scriptL_{sim}\rangle| \leq \ln
\scriptL_{sim}^{STD}$. Such models meet the necessary condition for
statistical compatibility, in the sense that the real lens sample may
actually be drawn from them. This regime is marked by the shaded area
in each panel of Fig.~6.

In conclusion, models in which early-type field galaxies form
homologously over a range of redshifts offer a vastly improved
representation of the actual lens sample.  Allowing some fraction of
galaxies to form at low redshift is an important component of these
models, though the mean star formation redshift is still relatively
high.  The best-fit models meet the necessary condition for
statistical consistency with the lens sample, based on the
interpretation of likelihoods. As noted in \S 3.1, altering the
assumed luminosity errors does not change this conclusion.

\section{Are Lensing Analyses Consistent?}

\subsection{Statistical Results}

The luminosity evolution of gravitational lens galaxies has recently
been explored by van de Ven et al.\ (2003), who use the FP to derive
an evolution rate of $d\log (M/L)_B/dz = -0.62 \pm 0.13$ from a sample
of 26 lenses. This is slightly faster than the findings of either R03
($-0.54 \pm 0.09$) or this paper ($-0.50 \pm 0.19$ for Model 1), even
though the lens samples and photometry are nearly the same.
Furthermore, vvF claim that rather late mean star formation redshifts
are permitted ($\langle z_f \rangle = 1.8^{+1.4}_{-0.5}$, 68\% C.L.),
while much of this range is excluded by R03 ($\langle z_f \rangle >
1.8$, 95\% C.L.)  and the combined structure/evolution analysis of \S
4.2.1 ($\langle z_f \rangle > 1.5$, 95\% C.L.). While the results of
RKK and vvF are technically consistent, it is instructive to discuss
three factors which account for the small differences: data weighting,
stellar evolution models, and sample selection.

First, differences in data weighting account for most of the
systematic disparity in evolution rates. We set a fixed uncertainty
for all galaxy luminosities, ensuring that each lens has a similar
relative contribution to the fit.\footnote{Recall that lenses with
estimated $z_s$ are weighted down relative to those with measured
$z_s$ (see \S 3.1).} In contrast, vvF use uncertainties which they
derive from their photometric conversion. Based on their quoted errors
on the surface brightness, some galaxies are far more important to the
fit than other galaxies. Data weighting has a surprisingly large
impact on the derived evolution rate. vvF already demonstrate this by
testing an error term (to account for scatter) which is added in
quadrature with other errors. This has the effect of weighting the
galaxies more uniformly, and slows the evolution rate from $d \log
(M/L)_B/dz = -0.62 \pm 0.13$ to $-0.56 \pm 0.12$. We have confirmed
these results by reanalyzing the data tabulated in vvF. Furthermore,
we find that the vvF data yield an even slower evolution rate of $d
\log (M/L)_B/dz = -0.46 \pm 0.12$ (very similar to our value) if we
impose uniform weighting.  The choice of weighting is clearly
important for constraining the evolution model, and the vvF method is
not unreasonable. However, we believe that our scheme provides a more
realistic description of the global properties of the lens galaxy
population, since uncertainties in Model 1 are certain to be dominated
by intrinsic scatter in the mass profile and formation redshift,
rather than by measurement errors. Galaxies should not be segregated
based on nominal photometric errors if those errors are much too small
to account for the observed scatter in the FP.  Moreover, weighting
galaxies by their photometric errors can introduce a Malmquist bias,
which we may be seeing in the vvF analysis.  Relatively brighter
galaxies are expected to have better photometry, giving them greater
statistical weight. Since brighter galaxies at high redshift will also
imply a greater degree of luminosity evolution, the mean evolution
rate will tend to be biased toward faster values. Nearby lens
galaxies, which are bright but have little ability to distinguish
between star formation models, would also be more strongly weighted
than the distant galaxies which are vital for determining stellar age
through luminosity evolution.

Second, the constraints on stellar age are affected by models of the
evolution tracks $e(z)$ as a function of star formation redshift. We use
the tracks derived from the BC96 stellar synthesis code. In contrast, vvF
use an approximation to the Worthey (1994) simulation results: $\log
(M/L)_B \propto \kappa \log (t - t_f)$, where $t-t_f$ is the stellar age
and $\kappa \simeq 0.93$. The models agree very well for $z_f \ga 2$, but
the BC96 model evolves more rapidly than the vvF fitting function for $z_f
\leq 1.5$. As a result, later star formation redshifts are more strongly
disfavored with the BC96 model, which tightens our lower bound on $\langle
z_f \rangle$. We tested this hypothesis by reanalyzing the vvF data (with
their quoted uncertainties). For example, while vvF find that their $z_f =
1.3$ track is marginally consistent with their data, the corresponding
BC96 track is unambiguously rejected at $\Delta \chi^2 = 12$. We believe
that the BC96 model facilitates more accurate constraints on the mean star
formation redshift, as it offers a more detailed treatment of young
stellar populations than the vvF approximation.

Third, differences in sample selection have a small systematic effect
on the evolution results. Systems without measured source redshifts
are rejected by vvF, but are included in our analysis (with an estimate of
$z_s = 2.0 \pm 1.0$). Two of these systems -- MG1131+0456 (Hewitt et
al.\ 1988) and B1938+666 (Patnaik et al.\ 1992) -- are particularly
interesting.  Each is at high redshift ($z_d=0.84$ and $0.88$,
respectively) and exhibits a larger than average mass-to-light ratio
(especially MG1131+0456). Hence, they force the model to slightly slower
evolution rates. While excluding such systems alters constraints on
the mean star formation redshift at the level of only $\sim 0.1$ (see
\S 4.2), there is no reason that these lenses should be rejected just
because they are outliers. In all other respects, they look like
typical elliptical galaxies (Tonry \& Kochanek 2000). One may argue,
however, that our choice for the estimated source redshift biases the
result. We note that the implied mass-to-light ratios of these
galaxies can be reduced if the sources are at a much higher redshift
($z_s \ga 4$). Based on the optical properties of the sources, this is
indeed a reasonable scenario. However, we find that these systems have
an equally important role in the fit even if we substitute
``measured'' values of $z_s = 4.0$ for our estimated values of $z_s =
2.0 \pm 1.0$ -- while the absolute deviation from the global model is
reduced, this is almost completely compensated by the reduced fit
tolerance for systems with a measured $z_s$. Therefore, the inclusion
of these systems in our analysis is strongly justified, and their
effects on the fit are real.

Finally, it is interesting to note that R03 and vvF have no systematic
differences in rest frame magnitudes, despite using alternate
techniques for photometric conversion. However, vvF appear to have
more scatter in their data points, and we trace this to the
photometry. Larger scatter slightly broadens the vvF confidence
regions compared to our own, and therefore leads to a somewhat weaker
rejection of late formation redshifts.

\subsection{Individual Lenses}

Several studies have focussed on the structure and evolution of
individual lens galaxies (Treu et al.\ 2002a,b; Koopmans \& Treu 2003;
Koopmans et al.\ 2003; Gebhardt et al.\ 2003; Treu \& Koopmans 2004).
While these analyses have the benefit of using measured velocity
dispersions in the fundamental plane relation, there are also two
significant drawbacks.  First, to determine the evolution implied by a
single galaxy or a small number of galaxies, one must make an explicit
comparison against a local FP intercept measured from other samples.
This often means using the intercept for ellipticals in the Coma
cluster to evaluate the evolution of field galaxies, even though these
populations should not, in general, have identical intercepts at
$z=0$. Second, such studies may be susceptible to a Malmquist bias,
which presumably makes it easier to measure velocity dispersions in
more luminous, higher surface brightness galaxies with younger stellar
populations. In this case, a sample of lenses with measured velocity
dispersions, especially at higher redshift, would be biased toward
faster evolution rates. Such effects may already be evident in the
current dynamical sample.

The Lens Structure and Dynamics survey (LSD; Koopmans \& Treu 2002,
2003; Treu \& Koopmans 2002a; Koopmans et al.\ 2003; Treu \& Koopmans
2004) has measured the velocity dispersions of several gravitational
lenses, which are combined with HST photometry to constrain structure
and evolution. Five of these lenses appear in our sample: MG2016+112
(Lawrence et al.\ 1984), 0047--281 (Warren et al.\ 1996), B1608+656
(Myers et al.\ 1995), HST14176+5226 (Ratnatunga et al.\ 1995), and
PG1115+080 (Weymann et al.\ 1980). As we discussed in \S 4.1, our
statistical constraints on the galaxy mass profile are largely
consistent with those derived through dynamical measurements and
modeling. Note, however, that Treu \& Koopmans (2004) now favor a mean
logarithmic slope which is significantly shallower and marginally
inconsistent with isothermal, based on a combined analysis of six lens
galaxies. Two of these galaxies do not currently appear in our sample,
and help weight the LSD analysis toward shallower profiles.

The LSD sample also favors faster evolution rates and younger stellar
populations than our results indicate. This may be due in part to a
luminosity-related selection bias, or at least small number
statistics. For example, our evolution results for MG2016+112 and
0047--281, based on a fixed local FP intercept, agree well with those
measured in the LSD survey -- but each lens scatters on the ``young''
side of our evolutionary trend-lines. Moreover, the LSD survey has
observed B1608+656, whose spectral properties suggest that the galaxy
has undergone significant star formation triggered by an ongoing
merger (Koopmans et al.\ 2003).  Using dust corrections based on
observed color gradients, they derive a luminosity which is much
larger than that of passively-evolving galaxies, and a luminosity
evolution which is much faster than our estimate for the overall
sample.\footnote{ The differences are exaggerated by the
foreground-screen extinction model used by Koopmans et al.\ (2003) to
correct the galaxy flux. A foreground screen produces the largest
correction, while a more realistic embedded screen or a mixture of
stars and dust would reduce the luminosity correction (e.g., Witt,
Thronson, \& Capuano 1992).  We note, however, that B1608+656 still
appears over-luminous in our model.}  Based on an analysis of their
lens sample, Treu \& Koopmans (2004) find an evolution rate of
$d\log(M/L)_B/dz = -0.72 \pm 0.10$, which is systematically faster but
still broadly consistent with the evolution rate we derive by
optimizing over the mass profile.

Departures from homology can also have interesting effects on the
derived evolution rates. If we compare the implied FP intercept with a
fixed local value, then a larger velocity dispersion yields a slower
evolution rate, and vice versa.\footnote{It is interesting to note
  that the evolution rate derived by Treu \& Koopmans (2004) slows
  dramatically if the local FP intercept is simultaneously fit.}
PG1115+080, whose measured velocity dispersion is much higher than the
isothermal prediction (Tonry 1998; Treu \& Koopmans 2002b), must
undergo virtually no luminosity evolution to $z=0.31$ if it is to fall
on the local FP. Alternatively, if the galaxy evolves at a standard
evolution rate, its structure cannot be consistent with the rest of
the early-type galaxy population.  HST14176+5226, in contrast, yields
a significantly faster evolution rate if the isothermal velocity
dispersion estimate is replaced with one of its measured values.
Unless the quoted error bars on the velocity dispersions have been
underestimated, the findings of Gebhardt et al.\ (2003), Treu \&
Koopmans (2004) and others suggest that early-type galaxies exhibit
significant structural diversity. Such deviations from homology warn
against using small numbers of lenses to constrain an evolution model
meant to describe the galaxy population as a whole.

\section{Summary and Discussion}

Evolution studies of early-type galaxies based on gravitational lens
samples (Kochanek et al.\ 2000; R03; vvF) have often been criticized
for using estimated, rather than measured, velocity dispersions,
because a mass model is required to convert the accurately measured
projected mass into a dynamical estimate. The primary concern has been
that the preferred mass model for lenses, an isothermal model with a
flat rotation curve, could be incorrect.  The secondary concern is
that early-type galaxies may not be well characterized by any single
mass distribution that is homologous to the luminosity distribution.
By simultaneously modeling both the structure and evolution of lens
galaxies, our present analysis helps address these concerns.
Specifically, we determine the best-fit mass profile for lens galaxies
(as well as the range over which individual lenses might scatter), and
account for these effects on the mean evolution model. Our technique
builds on the homologous two-component mass models of RKK, which were
constrained using the ensemble of aperture mass-radius relations
derived from lensed image geometries. The homology formalism is
virtually identical to the fundamental plane, with the AMR relations
substituting for measured velocity dispersions. In essence, this
allows us to trace luminosity evolution with the FP, while directly
using the quantity (aperture mass) that strong lensing naturally
constrains.

We first updated constraints on the structure of early-type galaxies,
assuming that lenses are characterized by a single star formation
redshift. We find that uncertainty in the evolution model has little
effect on our structural results. The FP strongly favors nearly
isothermal ($n=2$) mass profiles on scales of a few effective radii,
with a mean density slope of $n=2.06 \pm 0.17$ (68\% C.L.). These
statistical constraints are consistent with the isothermal paradigm
favored by modeling (Kochanek 1995; Cohn et al.\ 2001; Mu\~noz et al.\
2001; Rusin et al.\ 2002; Winn et al.\ 2003) and dynamical (Treu \&
Koopmans 2002a, 2002b; Koopmans \& Treu 2003; Koopmans et al.\ 2003)
studies of individual lens galaxies, the dynamics of local ellipticals
(e.g., Rix et al.\ 1997; Romanowsky \& Kochanek 2001; Gerhard et al.\
2001), and their X-ray halos (e.g., Fabbiano 1989; Matsushita et al.\
1998; Loewenstein \& White 1999).  Note, however, that Treu \&
Koopmans (2004) favor models with systematically shallower profiles,
while Romanowsky et al.\ (2003) favor significantly more concentrated
models.  Our analysis indicates that a small but non-zero fraction
($f_{cdm} > 0.08$, 95\% C.L.) of the mass projected inside two
effective radii must be in the form of an extended dark matter halo,
independent of any assumed stellar mass-to-light ratio.  We also
constrain the increase in the total mass-to-light ratio with galaxy
luminosity ($\Upsilon_* \propto L^x$) : $x=0.18^{+0.24}_{-0.18}$ for
the linear evolution model, and $x=0.14^{+0.22}_{-0.16}$ for the
detailed BC96 evolution tracks (both 68\% C.L.). These results are
consistent with recent dynamical analyses of early-type galaxies
(Gerhard et al.\ 2001; Bernardi et al.\ 2003; Borriello et al.\ 2003;
Padmanabhan et al.\ 2004).  Because the FP is closely related to the
homology formalism, our constraint on $x$ can be considered the first
constraint on the slopes of the lensing fundamental plane: $1.1 <
\alpha < 2.0$ and $0.3 < \beta < 0.4$. These constraints are not yet
very restrictive, but they are consistent with measurements of the
local FP (e.g., Jorgensen et al.\ 1996; Pahre et al.\ 1998a; Bernardi
et al.\ 2003).

Next we considered the luminosity evolution of lens galaxies, and find
a small effect from the uncertainties in the mass model. We began by
investigating a model in which all galaxies form at a common redshift
(Model 1). The same assumption has been made in almost every analysis
of early-type galaxies, and therefore provides a basis for comparing
results. The lens sample favors a linear evolution rate of $d \log
(M/L)_B / dz = -0.50 \pm 0.19$ (68\% C.L.). This constraint includes
the spread in structural properties that are consistent with the AMR
relations, and is therefore weaker than the value derived by R03 based
on isothermal profiles, $d \log (M/L)_B / dz = -0.54 \pm 0.09$ (68\%
C.L.). Assuming the standard cosmological parameters and a Salpeter
IMF, we require a mean star formation redshift of $\langle z_f \rangle
> 1.8$ at 68\% confidence, and $\langle z_f \rangle > 1.5$ at 95\%
confidence, based on the BC96 spectral population models.  These
constraints are remarkable, considering that we make no assumption
regarding the shape of the galaxy mass distribution or the slopes of
the fundamental plane. Our only assumption is that these galaxies are
structurally homologous, but even this is not restrictive: we obtain
identical conclusions if the increase in mass-to-light ratio with
luminosity is modeled as a systematic non-homology. We favor slightly
older stellar populations in lens galaxies than van de Ven et al.\
(2003), who suggest that lenses could be consistent with mean star
formation redshifts as late as $\langle z_f \rangle \sim 1.2$. We have
traced these differences to data weighting, evolution models and
sample selection.  In conclusion, our single-$z_f$ analysis again
indicates that, on average, early-type galaxies in low-density
environments have old stellar populations -- perhaps as old as their
counterparts in rich clusters.

Model 1 does not provide a good fit to the data given the estimated
statistical uncertainties, much as the scatter in the standard FP also
exceeds measurement errors. This suggests that additional sources of
complexity have not been properly taken into account. In particular,
we can be confident that galaxies are neither perfectly homologous nor
form at a common redshift.  The scatter about our best fitting
solution corresponds to a spread of approximately 30--35\% in some
combination of mass (deviations from homology) and luminosity (range
of star formation epochs).  For our overall constraints on the mean
structure and the mean star formation redshift, we included these
additional uncertainties in our error estimates.

We explored one source of this scatter using the seven lens systems in our
sample with direct velocity dispersion measurements. For each system
we can estimate the parameters of the mass model needed to reconcile
the velocity dispersion with the aperture mass constraint. In general,
all seven are consistent with our statistical results, given their
mutual errors.  This is just a rephrasing of the fact that our error
estimates in Model 1 have correctly included the scatter created by
deviations from homology or a spread in formation epochs. Four of the
seven lenses have parameters and parameter degeneracies similar to our
statistical models. PG1115+080 requires a steeper mass distribution,
while HST14176+5226 requires a shallower mass distribution.  However,
systematic errors in measuring and modeling stellar velocity
dispersions might significantly mitigate these discrepancies.
Q2237+0305, the bulge of a nearby spiral galaxy (Huchra et al.\ 1985),
should probably be described by a more complex mass model, but an
isothermal profile can still describe its inner dynamics.

We also investigated a model which can directly account for the
observed scatter in the homology relation by allowing a range of star
formation redshifts. Specifically, we assumed that stars form between
$z_{f,min}$ and $z_{f,max}$, with uniform probability density in $\log
z_f$, and that the galaxies are structurally homologous at formation.
By ignoring any intrinsic distribution of mass profiles, we expect to
maximize the range of formation redshifts necessary to reproduce the
observed scatter.  Applying a likelihood analysis to the lenses, we
find that a significant spread in formation redshift is favored:
($z_{f,max}-z_{f,min} > 0.4$ at 95\% confidence, and
$z_{f,max}-z_{f,min} > 1.7$ at 68\% confidence), while the
single-$z_f$ model is ruled out. Moreover, simulated data sets created from
a subset of the acceptable models yield likelihoods similar to that of
the real data.  Allowing a range of formation redshifts therefore
provides a vastly improved statistical description of the lens sample.
The required scatter in $z_f$ increases almost linearly with the mean
star formation redshift. This can be understood by noting that
luminosity evolution depends much more sensitively on $z_f$ when $z_f$
is small.  We find that the distribution of star formation redshifts
can vary significantly among acceptable models of the lens sample. For
example, the most favored models ($\Delta \scriptL < -1.15$) allow
anywhere from $0$ to $40\%$ of galaxies to form at $z_f< 1.5$
(Fig.~6). We therefore conclude that the stars in early-type field
galaxies form over a substantial range of redshifts, and that some
fraction of these stars may have formed as recently as $z_f\sim 1$.
Our analysis bolsters the findings of Treu et al.\ (2001, 2002), van
de Ven et al.\ (2003), Koopmans et al.\ (2003), and van der Wel et
al.\ (2004), who argue that late star formation in some fraction of
field ellipticals is essential for explaining their evolutionary
history. However, we depart from a few recent claims (Treu et al.\
2001, 2002; Gebhardt et al.\ 2003) that the mean star formation
redshift of field ellipticals may be as late as $z_f=1$.  In all of
our statistically acceptable models, the mean star formation redshift
is relatively high ($\langle z_f \rangle > 1.5$).

Understanding the star formation history of early-type field galaxies
remains a work in progress. There is, however, growing consistency
among results based on colors (Menanteau et al.\ 2001; vvF; Bell et
al.\ 2004), spectral properties (Bernardi et al.\ 1998; Schade et al.\
1999; Kuntschner et al.\ 2002; Treu et al.\ 2002) and the fundamental
plane (Kochanek et al.\ 2000; van Dokkum et al.\ 2001; Treu et al.\
2001, 2002; Rusin et al.\ 2003; van de Ven et al.\ 2003; van Dokkum \&
Ellis 2003; van der Wel et al.\ 2004; Treu \& Koopmans 2004; this
paper). Taken together, these analyses present a strong case that
field ellipticals formed the bulk of their stars between $z_f = 1.5$
and $2.0$, and are therefore not much younger than their counterparts
in high-density environments. This largely disagrees with the
predictions of semi-analytic CDM models (e.g., Baugh et al.\ 1996;
Kauffmann 1996; Kauffman \& Charlot 1998; Diaferio et al.\
2001). Moreover, based on the internal scatter of the various samples,
it appears that the stars in field ellipticals are likely to have
formed over a range of redshifts, with some fraction of the stars
forming as late as $z_f \sim 1$. The degree to which these galaxies
experience late star formation is a subject of continuing debate, and
several of the above results disagree on the details.

What is the source of these remaining discrepancies among analyses of
early-type field galaxies? Sample selection is probably the best
starting hypothesis.  Each of the above evolution studies employs
different criteria for selecting early-type galaxies. For example,
Treu et al.\ (2001, 2002) use morphological/magnitude/color cuts, van
Dokkum \& Ellis (2003) use morphological/magnitude cuts, and Gebhardt
et al.\ (2003) use the bulge fraction. The lens sample, by contrast,
is selected on mass, with large velocity dispersion ellipticals
dominating the lensing optical depth. The mass selection makes the
gravitational lens sample unique, as it is the only one not to be
affected by Malmquist biases related to luminosity, color, or other
star formation signatures.  However, lensing also selects more massive
samples of ellipticals.  Assuming an isothermal profile, the 22 lenses
we analyzed have a mean velocity dispersion of $\sim 240 \kms$, about
$10-25\%$ higher than any of the other samples listed above. The mean
velocity dispersion of lens galaxies also increases with redshift, due
to angular selection effects (R03). Because there is evidence to
indicate that more massive early-type galaxies may be older (van der
Wel et al.\ 2004), the higher mean velocity dispersion of the lens
sample could partially account for its slower evolution
rate. Additional work is clearly necessary to quantify the benefits
and biases of the various samples.

Whatever the remaining uncertainties, the lens galaxies are an
extraordinarily powerful probe of galactic structure and evolution.
Using only 22 lenses, we can simultaneously constrain the mean mass
profile of early-type galaxies, the dependence of the mass-to-light
ratio on luminosity, and the rate of stellar evolution, with
statistical and systematic uncertainties that are competitive with all
other methods. The power of gravitational lenses is a result of the
aperture mass-radius relations derived from the image geometries, but
good photometry and measured redshifts are needed to make use of these
constraints. Follow-up imaging and spectroscopy is therefore vital to
expanding the current lens sample, which may soon enable significantly
improved statistical analyses. Such future studies could investigate
lenses as a function of color, velocity dispersion, environment or
spectral properties. In addition, a sufficiently large sample of
lenses could probe not only luminosity evolution, but structural
evolution as well (e.g., Mao \& Kochanek 1994; Rix et al.\ 1994; Ofek,
Rix \& Maoz 2003), thereby tracing the merger history in low-density
environments. Detailed dynamical studies of individual lenses, such as
those being carried out by the LSD survey, are also vital, as they can
directly map the diversity of the lens galaxy population.  Similar
physical constraints can be obtained using time delay measurements in
gravitational lenses, if we can safely assume that the Hubble constant
is accurately measured.  With a known Hubble constant, time delays
measure the surface mass density near the lensed images (Kochanek
2002), which would break the model degeneracies for individual lenses
in the same manner as a velocity dispersion measurement.  Models for
the time variability created by the microlensing of lensed images can
also help break the degeneracy because the light curves can be used to
determine the fraction of the surface mass density in the form of
stars (Kochanek 2004; see also Schechter \& Wambsganss 2002 for a
discussion of statistical constraints based on flux ratio anomalies).
Because of their unique properties, gravitational lenses will continue
to contribute significantly to our knowledge of early-type galaxies,
and provide a testing ground for galaxy formation theories.

\acknowledgements

We thank the anonymous referee for offering suggestions which greatly
improved the original manuscript. We acknowledge the support of HST
grants GO-7495, 7887, 8175, 8804, 9133 and 9744. We acknowledge the
support of the Smithsonian Institution. CSK is supported by NASA ATP
Grant NAG5-9265. This work is based on observations made with the
NASA/ESA Hubble Space Telescope, obtained at the Space Telescope
Science Institute, which is operated by AURA, Inc., under NASA
contract NAS 5-26555.

\begin{figure*}
\psfig{file=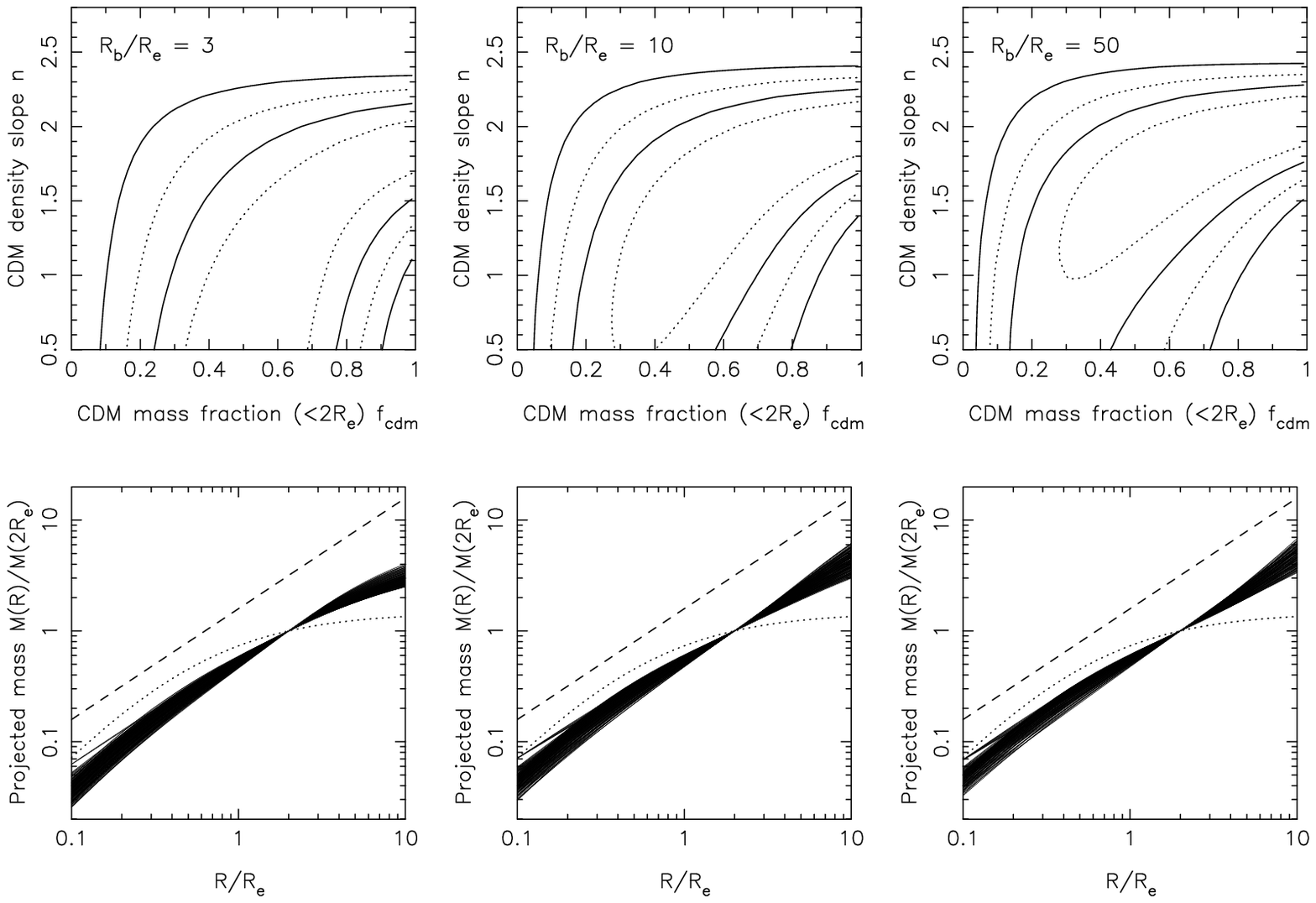}
\figurenum{1}
\caption{
  The structure of lens galaxies. All calculations assume a single
  star formation redshift (Model 1), but there is little change if we
  allow for a range of star formation redshifts (Model 2). Top:
  Constraints on the shape of the mass profile, which is set by the
  CDM mass fraction inside 2 effective radii ($f_{cdm}$), and the
  inner logarithmic slope of the CDM halo ($n$, with density $\rho
  \propto r^{-n}$). The halo is modeled with a cuspy profile (Mu\~noz
  et al.\ 2001), which breaks to an outer logarithmic slope of $3$ at
  a characteristic radius $R_b$.  We show constraints for models with
  $R_b/R_e = 3$, $10$ and $50$, where $R_e$ is the optical effective
  radius. Solid contours represent $\Delta \chi^2 = 2.30$ and $6.17$,
  the 68\% and 95\% confidence levels for two parameters.  Dotted
  lines represent $\Delta \chi^2 = 1$ and $4$, the 68\% and 95\%
  confidence levels for one parameter. The errors have been rescaled
  so that the best-fit model has $\chi^2 = N_{DOF}$.  Bottom: The mass
  profiles of models favored at 68\% confidence.  Solid lines are the
  projected masses inside $R/R_e$. Profiles are normalized to a fixed
  projected mass at $R=2R_e$. For comparison we show the de
  Vaucouleurs profile (dotted line), and an offset isothermal profile
  (dashed line). While the allowed models exhibit a wide range of dark
  matter abundances and break radii, they all have total mass profiles
  which are approximately isothermal over the radial range spanned by
  the lensed images.}

\end{figure*} 

\clearpage
\begin{figure*}
\psfig{file=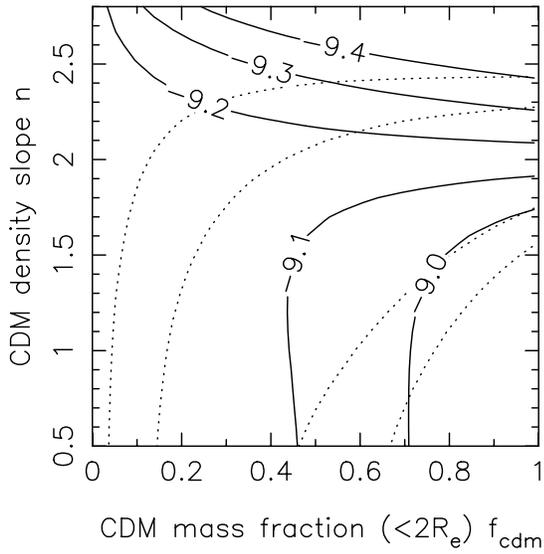,width=2.8in} 
\figurenum{2}
\caption{The recovered local FP intercept $\gamma_0$ as a function of
  mass profile, for $h=0.65$. We assume isotropic orbits ($\beta_{iso}
  = 0$) and a cuspy dark matter halo with $R_b/R_e = 50$. Velocity
  dispersions are estimated in an aperture with radius $R_e/8$. The FP
  slopes are fixed at $\alpha = 1.25$ and $\beta = 0.325$,
  corresponding to $x=0.3$. Solid lines are contours of best-fit
  $\gamma_0$. Dotted lines are the 68\% and 95\% confidence limits on
  the model (in two dimensions). These limits are slightly different
  than those shown in Fig.~1 because we have fixed the value of $x$.
  We find that the local FP intercept $\gamma_0$ varies slowly over
  our parameter space, and must therefore be measured to a precision
  of $\ll 0.1$ to improve the model constraints. Bender et al.\ 
  (1998) measures $\gamma_0 \sim -9.0$ for $\alpha=1.25$ and $\beta =
  0.32$. }

\end{figure*}

\clearpage
\begin{figure*}
\psfig{file=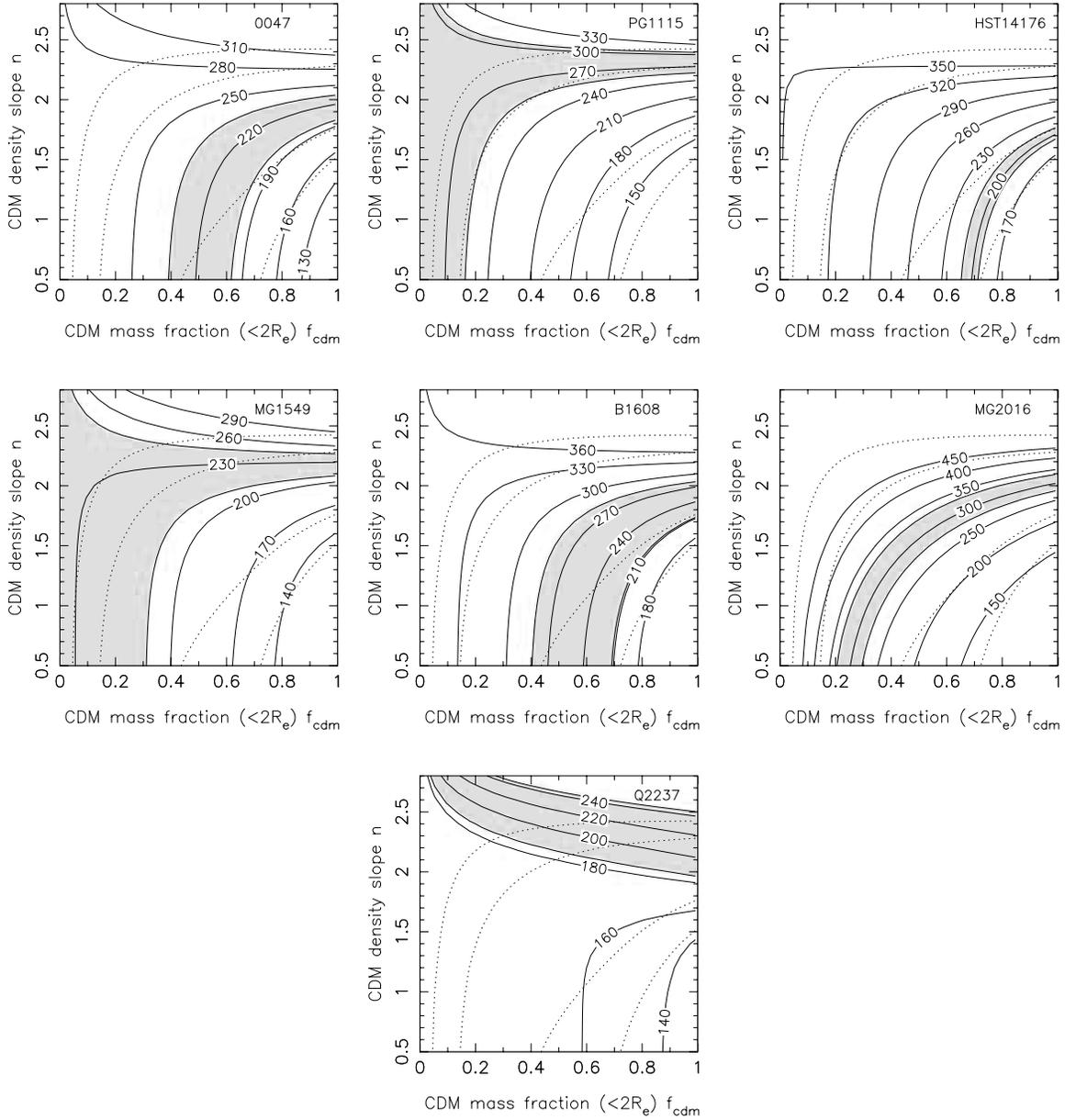,width=6in}
\figurenum{3}
\caption{Velocity dispersions for lens galaxies. Solid lines are
  contours of constant velocity dispersion, estimated in circular
  apertures with area equal to that of the respective apertures in
  which they were observed. Dotted lines are the two-dimensional model
  constraints for $R_b/R_e = 50$, shown in Fig.~1. The shaded areas
  are the measured velocity dispersions (68\% C.L.). As previously
  noted, the dynamics of two galaxies may require profiles which are
  significantly different than isothermal. PG1115+080 is steeper,
  while HST14176+5226 is shallower. Substituting the higher Ohyama et
  al.\ (2002) or Treu \& Koopmans (2004) velocity dispersion for the
  Gebhardt et al.\ (2003) value would bring HST14176+5226 into better
  agreement with our statistical results.}
\end{figure*}

\clearpage
\begin{figure*}
\psfig{file=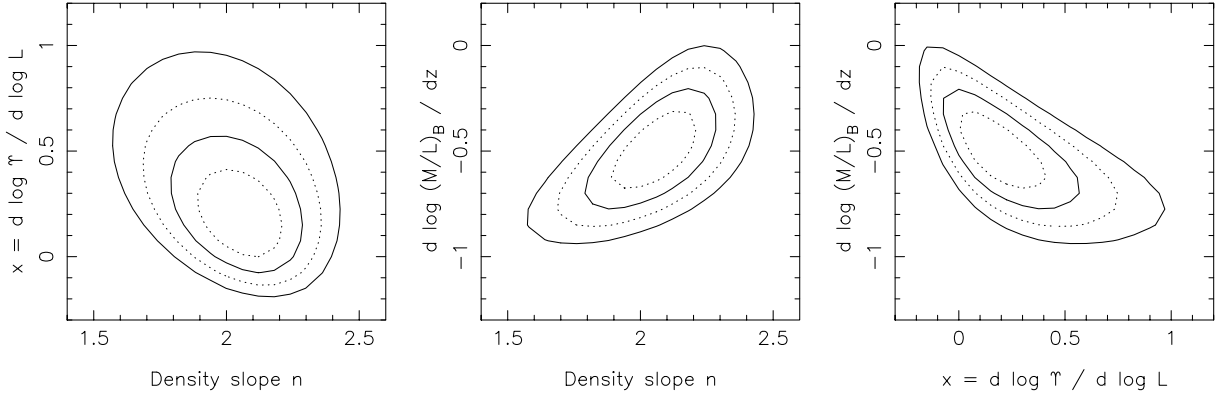}
\figurenum{4}
\caption{
  Constraints on the structure and evolution rate in Model 1. Plotted
  are pairwise constraints on the linear evolution rate $d \log
  (M/L)_B/dz$, mass-to-light ratio index $x$, and logarithmic density
  slope $n$ for a scale-free total mass distribution. Contours are
  drawn as in Fig.~1.}
\end{figure*}

\begin{figure*}
\psfig{file=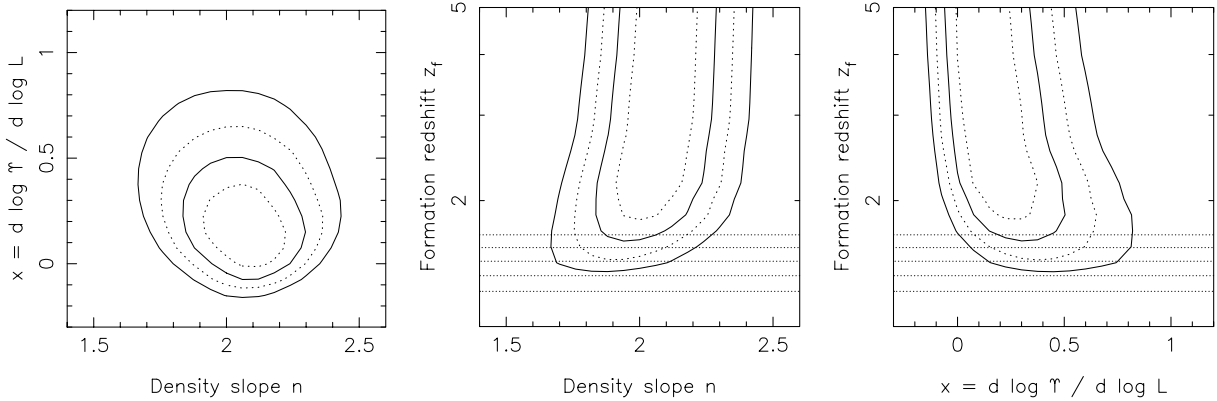}
\figurenum{5}
\caption{Constraints on the structure and mean star formation redshift
$\langle z_f \rangle$ in Model 1. The panels are analogous to those in
Fig.~4. Horizontal dotted lines mark $\langle z_f \rangle = 1.3$,
$1.4... 1.7$. The lens sample favors old stellar populations, with
$\langle z_f \rangle > 1.5$ at 95\% confidence.}
\end{figure*}

\clearpage
\begin{figure*}
\psfig{file=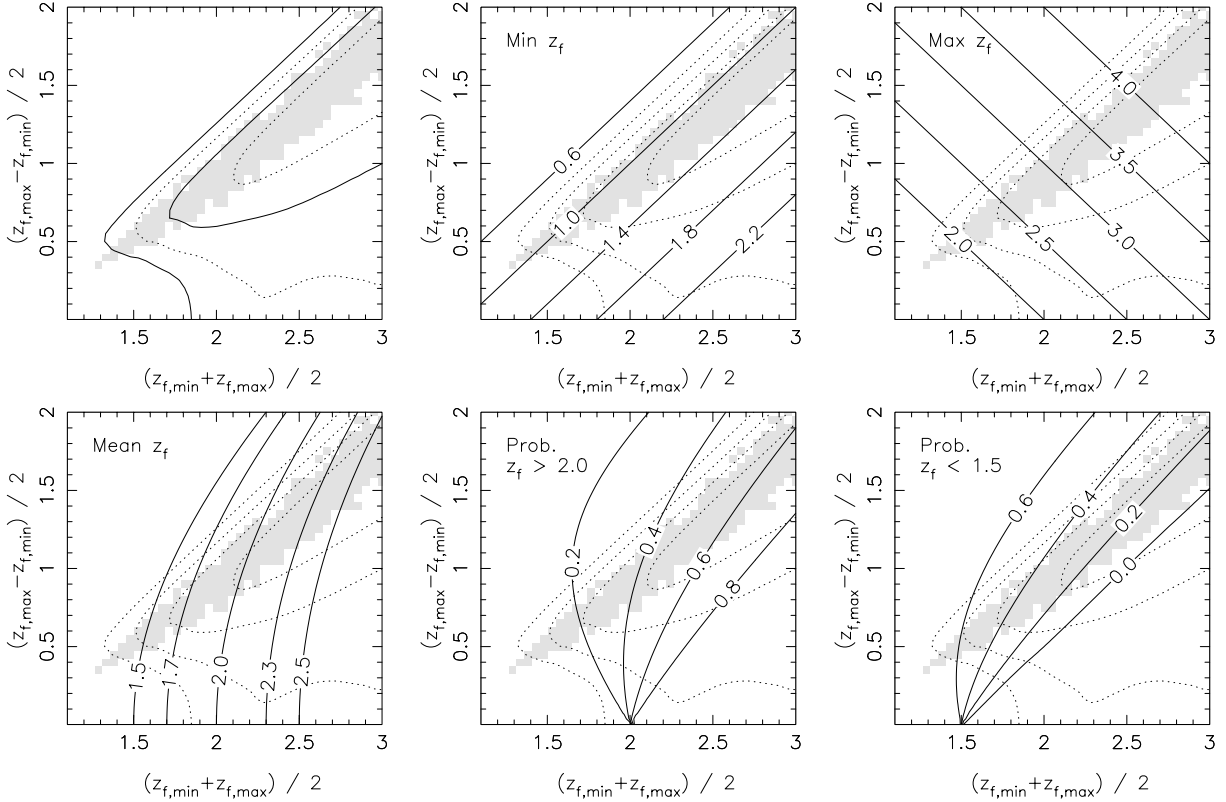}
\figurenum{6}
\caption{ Top left: Constraints on the detailed star formation history
  of lens galaxies (Model 2). Stars form over an interval spanning
  $z_{f,min}$ to $z_{f,max}$, with uniform probability density in
  $\log z_f$.  Confidence limits are plotted using solid and dotted
  contours as before, and are calculated from likelihood
  differences. We have optimized over all other model
  parameters. Significant scatter in $z_f$ is favored, and the
  single-$z_f$ scenario is excluded. Models which match the
  statistical properties of the actual lens sample are shaded in each
  panel.  For these models, the lens data are consistent with the
  likelihood distribution determined from simulated data sets. The
  remaining panels show contours illustrating the distribution of star
  formation redshifts. Here our previous constraints are shown as
  dotted contours.  Top center: Minimum $z_f$. Top right: Maximum
  $z_f$. Bottom left: Mean $z_f$. Bottom center: Fraction of galaxies
  forming at $z_f > 2.0$. Bottom right: Fraction of galaxies forming
  at $z_f < 1.5$.}
\end{figure*}

\end{document}